%% file: main.tex
\documentclass[conference]{IEEEtran}
\IEEEoverridecommandlockouts
\usepackage{cite}
\usepackage{amsmath,amssymb,amsfonts}
\usepackage{algorithmic}
\usepackage{graphicx}
\usepackage{textcomp}
\usepackage{xcolor}
\def\BibTeX{{\rm B\kern-.05em{\sc i\kern-.025em b}\kern-.08em
    T\kern-.1667em\lower.7ex\hbox{E}\kern-.125emX}}
\usepackage{soul}
\usepackage{subfig}

\usepackage{multirow}
\usepackage{booktabs}

\usepackage{array} 
\usepackage{hyperref}
\hypersetup{
    colorlinks=true, 
    allcolors=blue,
}
\usepackage{xurl}

\usepackage{alphalph}

\usepackage{rotating}

\input{macro}

\begin{document}

\title{The Anatomy of Silent Data Corruption:\\GPU Error Pattern Study and Modeling Guidance}

\author{
Chung-Hsuan Tung$^{1,*}$\thanks{$^{*}$This work was performed when authors were affiliated with NVIDIA.},
Yanxiang Huang$^{2}$,
Nirmal Saxena$^{2}$,
Philip Shirvani$^{2}$\\
Saurabh Hukerikar$^{2}$,
Twinkle Jain$^{2}$,
Abhishek Tyagi$^{3,*}$,
Sanjay Gongalore$^{2}$\\[0.5ex]
$^{1}$Duke University, $^{2}$NVIDIA, $^{3}$University of Rochester\vspace{-3mm}}

\maketitle


\begin{abstract}


Silent data corruption (SDC) threatens the reliability of large-scale GPU clusters used for training large language models, yet its rarity and lack of explicit error signals make accurate high-level modeling challenging.
To address this gap, we conducted a large-scale gate-level stuck-at fault injection on a production-class data-center GPU, consuming over three million simulator hours across 63 CUDA micro-benchmarks.
We extracted GPU SDC characteristics in terms of corruption types, bit-flip behavior, and warp-aligned spatial correlation.
Our results show that NaN/$\pm$INF account for only 1.01\% of SDC outcomes, that single-bit flips constitute less than 40\% of bit-flip events, and that corruption addresses exhibit periodicity.
These statistics motivate distribution-aware high-level fault modeling and realistic software-based fault injection for resilience evaluation of production-class GPU architectures.

\end{abstract}


\begin{IEEEkeywords}
Silent Data Corruption (SDC), Graphics Processing Unit (GPU), Fault Injection (FI), Hardware Reliability
\end{IEEEkeywords}

\input{1_intro}
\input{2_fi}

\input{3_pattern}

\input{4_exp_setup}

\input{5_results}

\input{6_impact}
\input{7_related_work}

\input{8_conclusion}



\bibliographystyle{ieeetr}
\bibliography{reference}

\end{document}

%% file: macro.tex
\newcommand{\myparatight}[1]{\vspace{0.5ex}\noindent\textbf{#1~}}

\newcounter{myObservationCounter}
\newcommand{\myobservation}[1]{
  \stepcounter{myObservationCounter}
  \vspace{0.5ex}\noindent\textbf{Observation \arabic{myObservationCounter}.~}\textit{#1}\vspace{0.5ex}
}

\newcounter{myInsightCounter}
\newcommand{\myinsight}[1]{
  \stepcounter{myInsightCounter}
  \vspace{0.5ex}\noindent\textbf{Insight \arabic{myInsightCounter}.~}\textit{#1}\vspace{0.5ex}
}

\newcolumntype{+}{>{\global\let\currentrowstyle\relax}}
\newcolumntype{^}{>{\currentrowstyle}}

%% file: 1_intro.tex
\section{Introduction}




Silent data corruption (SDC) is a \emph{symptom} of hardware faults in which a processing unit produces incorrect results without any hardware indication~\cite{sdc_whitepaper_ocp, hochschild2021cores, dixit2021silent}.
Open Compute Project (OCP)~\cite{sdc_whitepaper_ocp} defines outcomes of hardware faults as \emph{benign} if they leave program output unaffected, \emph{corrected} if detected and repaired by protections (e.g., error correction code (ECC) in memories), \emph{detected unrecoverable errors (DUEs)} if flagged by error detectors (e.g., parity and instruction validation) but the output cannot be corrected, and SDCs.
SDCs are considered the most insidious outcomes, as they escape internal detections when corrupting program output~\cite{sdc_whitepaper_ocp}.


Modern machine learning (ML) algorithms, especially large language models (LLMs), have shifted computation to massive GPU clusters where even rare events become non-negligible.
Particularly, training is resource-intensive: GPT-3 with {175}\thinspace{B} parameters~\cite{brown2020gpt3} required nearly 300 GPU-years on NVIDIA V100s~\cite{narayanan2021megatron}, and Llama 3 with {405}\thinspace{B} parameters used up to {16}\thinspace{K} NVIDIA H100 GPUs~\cite{grattafiori2024llama}.
SDCs have been observed at scale by Google~\cite{hochschild2021cores}, Meta~\cite{dixit2021silent}, and Alibaba~\cite{wang2023understanding}, causing 1.4\% of unexpected GPU interruptions in Llama 3 training~\cite{grattafiori2024llama}, and reported weekly/biweekly in Gemini training~\cite{team2023gemini}.

Despite robust hardware protection mechanisms, faults in unprotected logic can still lead to SDC~\cite{hochschild2021cores, he2023understanding}, which becomes especially harmful for modern LLMs that scale across many GPUs.
Insufficient understanding of SDC characteristics and impacts leads to high handling costs~\cite{hochschild2021cores, dixit2021silent}, i.e., detection requires dual modular redundancy with comparison (2$\times$ overhead), and correction needs triple modular redundancy (3$\times$ overhead)~\cite{fiala2012detection}.
Suspicion-based detections~\cite{comanici2025gemini, hochschild2021cores} monitoring anomalies, e.g., not-a-number (NaN), infinities ($\pm$INF), or the consequent loss spike~\cite{he2023understanding, ma2025understanding}, also suffer from human-centric debugging~\cite{dixit2021silent}, and long-range, unreliable rollback~\cite{comanici2025gemini} due to delayed degradation of ML training accuracy~\cite{he2023understanding}.

Fault injection (FI) can evaluate SDC impact and guide mitigation. While FI is standard in accelerator design~\cite{tung2023dynamic} and automotive functional safety~\cite{lotfi2019resiliency, hukerikar2024optimizing}, high-fidelity FI depends on the fault model, stimuli, and hardware details.

Prior work either uses software-level injectors with hypothetical data corruption patterns~\cite{bolchini2024resilience, tsai2021nvbitfi, mahmoud2020pytorchfi, chen2020tensorfi}, reports CPU-cluster observations~\cite{dixit2021silent, wang2023understanding, hochschild2021cores}, or studies reduced-complexity accelerators~\cite{he2023understanding, he2020fidelity}.
Consequently, corruption patterns in production-class data-center GPUs remain underexplored.

This paper addresses that gap by characterizing structural SDC manifestations and spatial correlations via gate-level FI on a production-class GPU architecture, with the goal of informing higher-level modeling and mitigation strategies.
Our campaign yields the following contributions:
\begin{itemize}
\item A statistical characterization of GPU SDC manifestations, providing distributions over corruption types, multi-bit flip patterns, and warp-aligned spatial correlation.
\item A large-scale (three million simulator-hours) gate-level FI campaign on a production-class data-center GPU model.
\item A CUDA micro-benchmark suite for systematic structural fault characterization across GPU functional units.
\end{itemize}



Our campaign clarifies the structural nature of GPU SDC, supporting realistic high-level fault modeling and injection.


%% file: 2_fi.tex
\section{Fault Injection and High-level Modeling}


\begin{figure}[!t]
    \includegraphics[width=0.98\columnwidth]{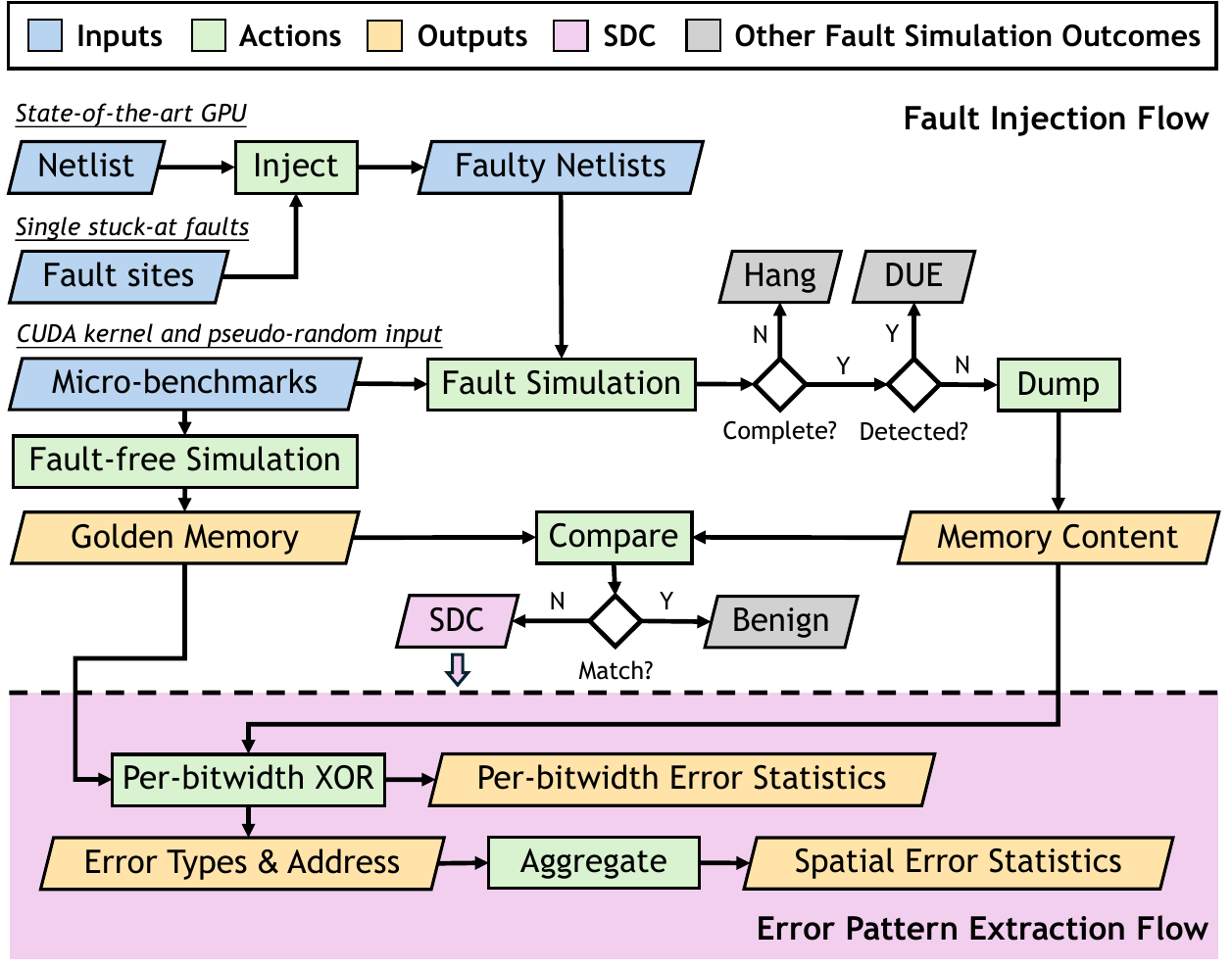}
    \caption{Fault injection and error pattern extraction flow. We analyze the corrupted memory to characterize the SDC behavior.}
    \vspace{-3mm}
    \label{fig: extraction_flow}
\end{figure}

\subsection{Gate-level Fault Injection}


FI consumes the hardware design netlist, a fault model, and test vectors, as shown at the top left of Fig.~\ref{fig: extraction_flow}.
We craft micro-benchmarks (Sec.~\ref{subsec: ubench}) as test vectors to systematically exercise critical GPU data paths.
For each benchmark, injecting a single fault yields one of four outcomes: hang, benign, DUE, or SDC.
For the SDC outcome, the simulation dumps the corrupted memory, which is compared to the golden memory from the fault-free simulation for pattern analysis.



\myparatight{Fault Model.}
Our study targets permanent faults arising from physical defects, such as wear-out and process variation~\cite{dixit2021silent}. Unlike transient faults studied in~\cite{he2020fidelity}, which can be abstracted as memory bit flips, permanent faults in combinational logic require specific modeling due to their persistent impact.

To enable large-scale FI, we utilize the single stuck-at fault (SAF) model. While advanced models like bridging, small delay~\cite{tehranipoor2011test}, cell-aware~\cite{hapke2014cell}, and physically-aware region faults~\cite{li2022pepr} offer higher accuracy, they are generally unsupported by the simulation engines of commercial EDA tools.

Our reliance on the SAF model is justified by the fact that the emergent error patterns, specifically bit-flip distributions and spatial correlations, are largely a function of the logic network's internal topology. Most localized faults (including bridging/delay defects) propagate through the same sensitized paths to the primary outputs, yielding similar error signatures. Cell-aware or physically-aware region faults may result in more expansive error patterns due to wider internal fan-out, but the core topological propagation remains consistent.

\myparatight{Simulation Outcomes.}
Faults manifest as \emph{hang}, \emph{benign (masked)}, \emph{DUE}, or \emph{SDC}~\cite{saxena2022error, tsai2021nvbitfi}.
\emph{hang} means the runtime exceeds twice the fault-free execution;
\emph{DUE} means a machine check exception is observed.
Otherwise, memory contents are compared to the golden run: identical marks the fault as \emph{benign} and \emph{masked}; a mismatch implies an \emph{SDC}.
Simulations in which the injected fault causes an SDC are studied to understand the spatial signature of the SDC in the program output (Sec.~\ref{sec: pattern_analysis}).

\subsection{Micro-benchmarks}\label{subsec: ubench}

A suite of 63 micro-benchmarks exercises critical GPU pipelines through combinations of functional units, operations, data types, and input stimuli. 
Each benchmark pairs a CUDA kernel (an operation with a specific data type executed on the ALU or tensor core) with input sequences generated by Mersenne--Twister (MT)~\cite{matsumoto1998mersenne}, universal test pattern (UTP)~\cite{fujiwara1981design, fujiwara1984new, shen1984design, hukerikar2022runtime}, or linear feedback shift register (LFSR)~\cite{abramovici1990digital, golomb2017shift}.
Each operand uses a distinct sequence under the selected method.
Operations include addition, multiplication, fused multiply-add (FMA), and general matrix multiplication (GEMM, $\alpha A B + \beta C$).
We evaluate GEMM in addition ($B=I$, where $I$ is identity matrix), multiplication ($\beta=0$), and standard ($\alpha=\beta=1$) form.
Data types span UINT8/UINT32 and FP16/FP32/BF16/TF32, with limited exploration of FP8.




%% file: 3_pattern.tex
\section{Memory Corruption Pattern Analysis Method}\label{sec: pattern_analysis}





We analyze the \emph{locations} (memory addresses) and \emph{values} of corruptions in dumped memory images.
As software observes the memory state, per-kernel statistics from the corresponding micro-benchmark can be applied to higher-level modeling.


\myparatight{Construction of Error Profile.}
Fig.~\ref{fig: extraction_flow} shows the detailed FI flow (top) with the error pattern extraction in the case of SDC (bottom).
The FI experiments will dump the raw memory image, which stores the program outputs.
%
%
We extract the bit-level error pattern from each dumped image and its corresponding golden memory using bitwise XOR for every bit-width, e.g., 32-bit for FP32.
We record the list of differing addresses and values, compute per-benchmark corruption rates, bucket values into error types, and accumulate bit flip probabilities.
Addresses are analyzed for spatial patterns for each error type.


\myparatight{Corruption Value Patterns.}
We classify corruptions as \emph{nullified} (all bits zero) or \emph{bit-flipped}.
Nullification is common when SAFs in control logic disable computations or select data paths incorrectly.
For bit flips, we track flip rates per bit position and the distribution over flip counts in a bit-width.
For floating-point types, we separately record rates for special values (NaN/$\pm$INF) that are often targeted by software checks.

\myparatight{Corruption Location Patterns.}
%
Grouping corrupted addresses modulo the warp size ($W{=}32$) reveals periodic corruption patterns at the warp granularity.
While the exact stride may depend on the GPU model, the observed spatial correlation suggests structured rather than random fault propagation.

%% file: 4_exp_setup.tex
\section{Experimental Setup}



\myparatight{GPU Architecture.}
%
%
We use a synthesized gate-level model of a production-class data-center GPU in a reduced two-SM configuration. 
Each SM integrates scalar/tensor pipelines, warp scheduling, shared memory, and a private L1 cache, and connects via an on-chip network to a shared L2 and high-bandwidth memory.
The model executes CUDA-derived traces under a 32-thread SIMT execution model with an ECC-protected cache and memory hierarchy.
Product identifiers are omitted; however, the modeled architecture reflects that of a deployed data-center GPU used for ML workloads.



\myparatight{Fault Sampling.}
Using a commercial EDA tool, we extract SAF sites with hierarchical net names to enable hardware-unit attribution and uniformly sample them across the full SM logic, while ensuring each failure mode is adequately represented~\cite{hukerikar2024optimizing}.
Stuck-at-0/1 are randomly assigned, resulting in an approximately balanced distribution.
Given the high masked rates~\cite{saxena2022error, hochschild2021cores, dixit2021silent} and cost of gate-level FI, we apply a testability check to ensure stimulus reachability.
For a representative micro-benchmark, this filtering reduces the candidate sites from 6{,}925 to 1{,}507 (78.2\%), with similar reductions observed across other micro-benchmarks.


\myparatight{Memory Image.}
Micro-benchmarks are structured to have a single contiguous output memory space.
We track those memory spaces and simulate all benchmark--fault combinations.

\myparatight{Output Data Type.}
Although benchmarks test data paths of UINT8/UINT32 and FP8/FP16/FP32/BF16/TF32, the memory images are limited to UINT32 and FP8/FP16/FP32.
This is because UINT8 and BF16 are supported only on tensor units, whose accumulators store in UINT32 and FP32, respectively; TF32 is stored in FP32 by design.

%% file: 5_results.tex
\section{SDC Pattern Analysis and Observations}\label{sec: results}

%
%
\begin{figure}[!t]
    \centering
    \subfloat[Corruption counts normalized by memory size]{\includegraphics[width=0.98\columnwidth]{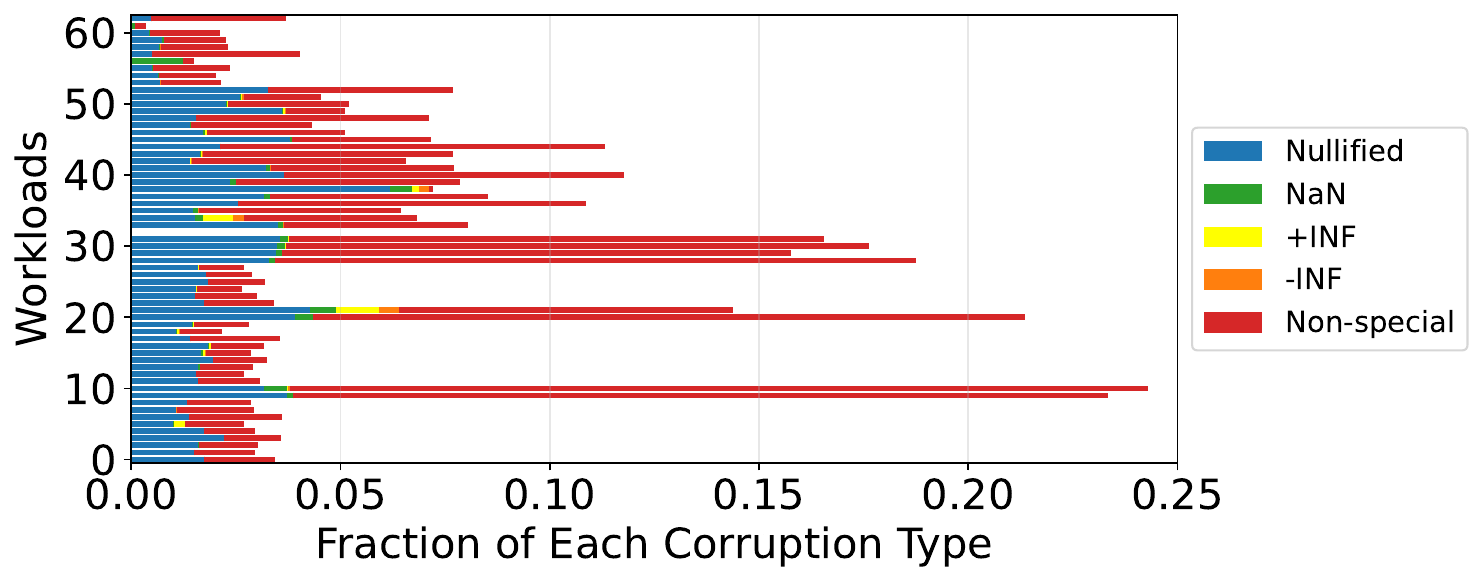}\label{subfig: corrupt_type_norm_surface}}
    \vspace{0mm}
    \subfloat[Normalized corruption type breakdown per benchmark]{\includegraphics[width=0.98\columnwidth]{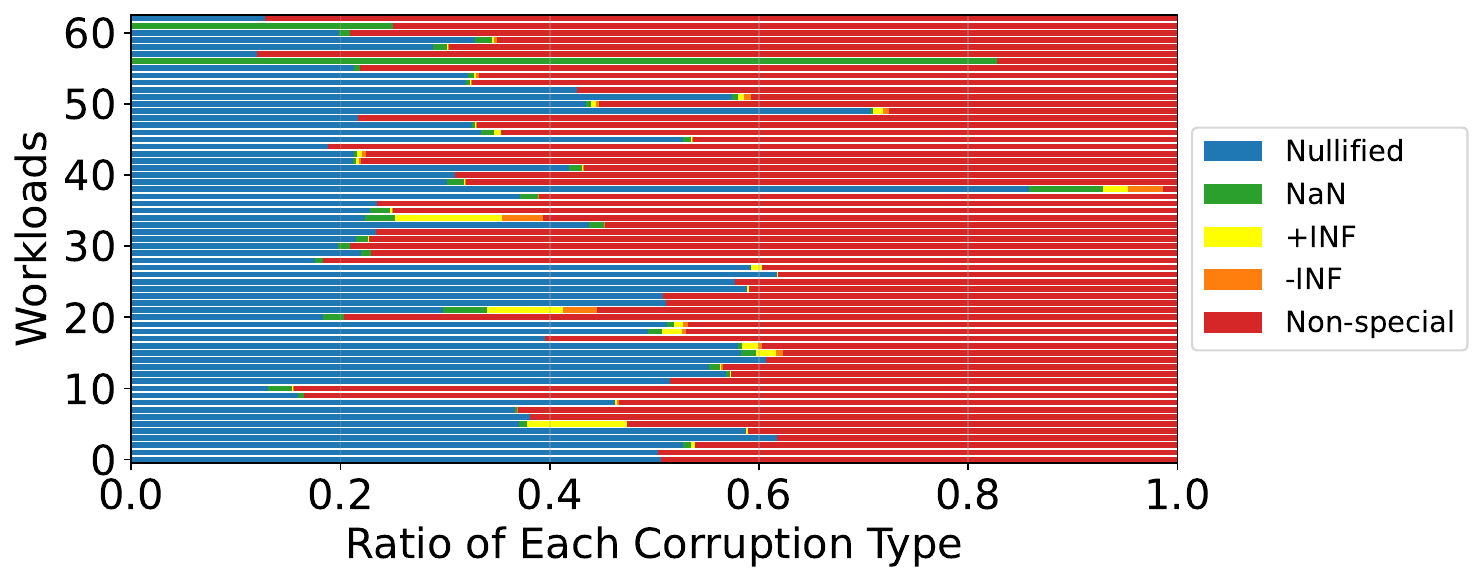}\label{subfig: corrupt_type_norm_1}}
    \caption{Corruption types across 63 micro-benchmarks. NaN, +INF, -INF, and Non-special are all bit flips; software-detectable special values are rare relative to other corruptions.}
    \vspace{-3mm}
    \label{fig: corrupt_all_units}
\end{figure}




Across all benchmark--fault pairs, we observe {25k} SDC cases and {28k} DUEs or hang outcomes. 
For each SDC case, we analyze the output memory image ({1.8}\thinspace{KB}--{64}\thinspace{MB}), and report aggregate statistics with key observations.


%

\subsection{Corruption Outcome Distribution}


Fig.~\ref{fig: corrupt_all_units} summarizes corruption rates regarding the memory space in Fig.~\ref{fig: corrupt_all_units}\subref{subfig: corrupt_type_norm_surface} and type breakdowns in Fig.~\ref{fig: corrupt_all_units}\subref{subfig: corrupt_type_norm_1}.

\myobservation{Special values (NaN and $\pm$INF) are rare.}

Across 600\thinspace{M} observed corruptions, NaN/+INF/-INF account for 0.39\%/0.49\%/0.13\%, respectively, i.e., 1.01\% jointly.
This observation reveals that the efficacy of software-based SDC screening of NaN/$\pm$INF is potentially limited to a small subset of SDCs.
The rarity of NaN/$\pm$INF arises because producing these values requires flipping the entire exponent field to ones, an event with a large Hamming distance from typical values.
The exponent logic is also more reliable due to its simplicity compared to mantissa logic (Sec.~\ref{subsec: bitflip_stats}).





\myobservation{Nullification is prevalent.}

Despite balanced stuck-at-0/1 injections, corrupted outputs are strongly zero-biased: nullification accounts for 50.68\% of corruptions, making it 50.17$\times$ more likely than NaN/$\pm$INF.
This dominance is consistent with control-path SAFs that disable computation or disturb addressing, thereby preserving initialized zero states.
Given its prevalence and distinct semantics, we separate nullification from other corrupt-but-valid values (48.31\%).
The latter category (non-special bit-flips) is further analyzed at the bit level in Sec.~\ref{subsec: bitflip_stats}.

\begin{figure}[!t]
    \centering
    \includegraphics[width=0.98\columnwidth]{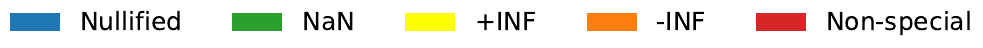}
    \vspace{-3mm}

    \subfloat[CudaCoreControl1]{
        \includegraphics[width=0.46\columnwidth]{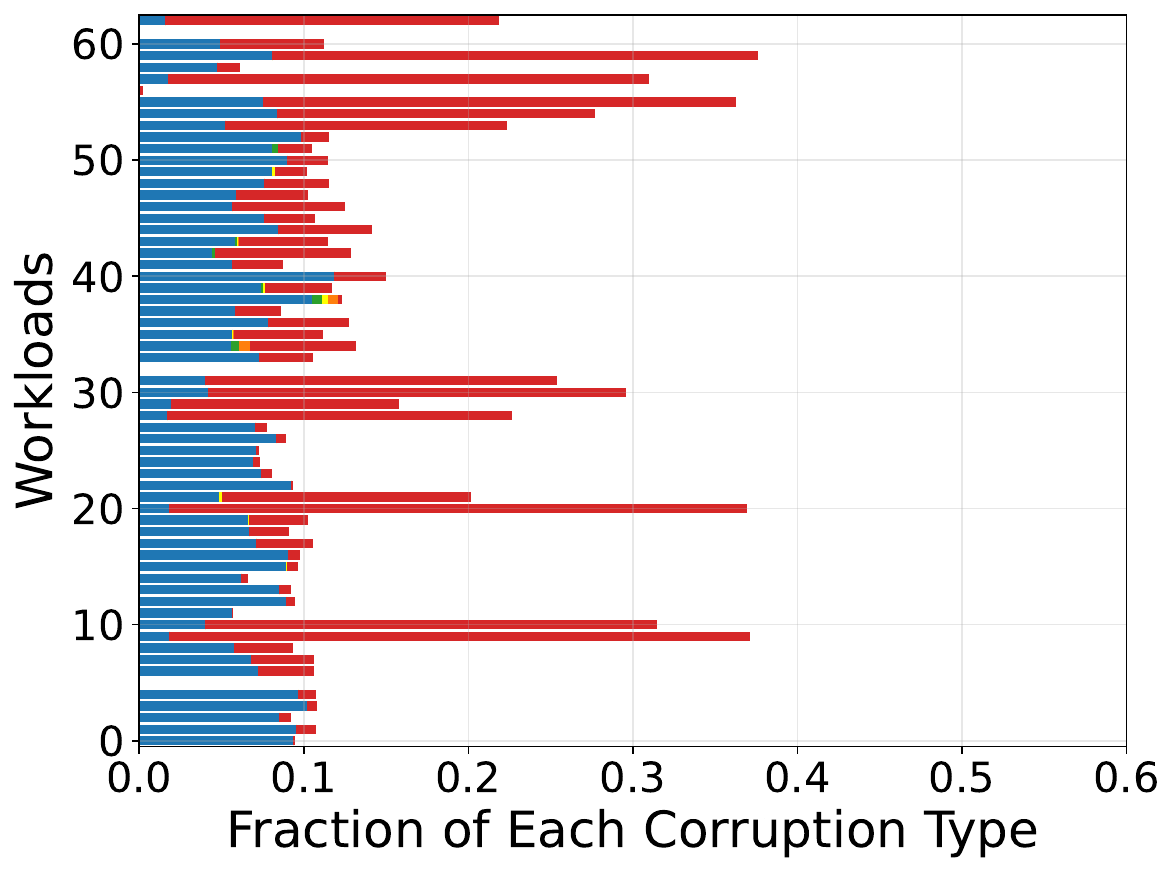}
        \label{subfig: norm_surface_sctl}}
    \vspace{-5mm}
    \subfloat[CudaCoreControl2]{
        \includegraphics[width=0.46\columnwidth]{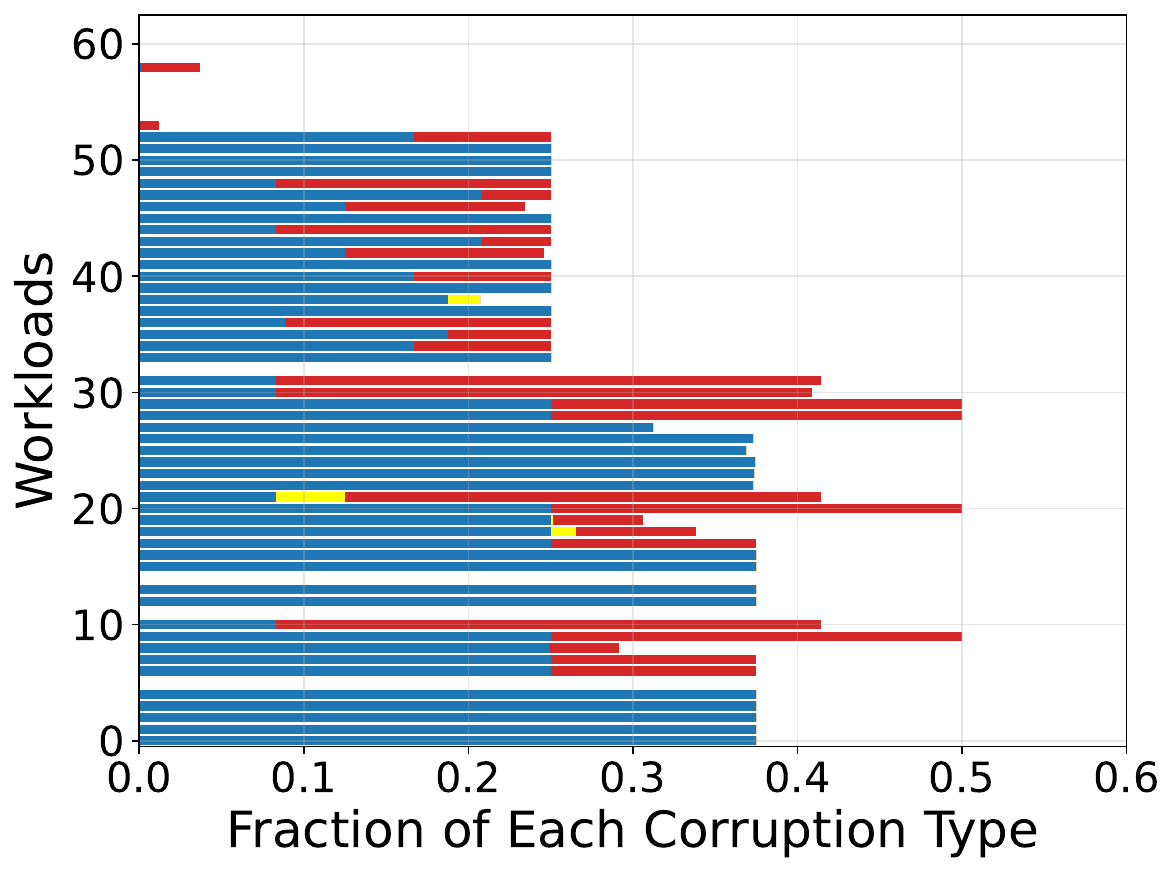}
        \label{subfig: norm_surface_sfe}}
    \vspace{2mm}
    \subfloat[TensorCore]{
        \includegraphics[width=0.46\columnwidth]{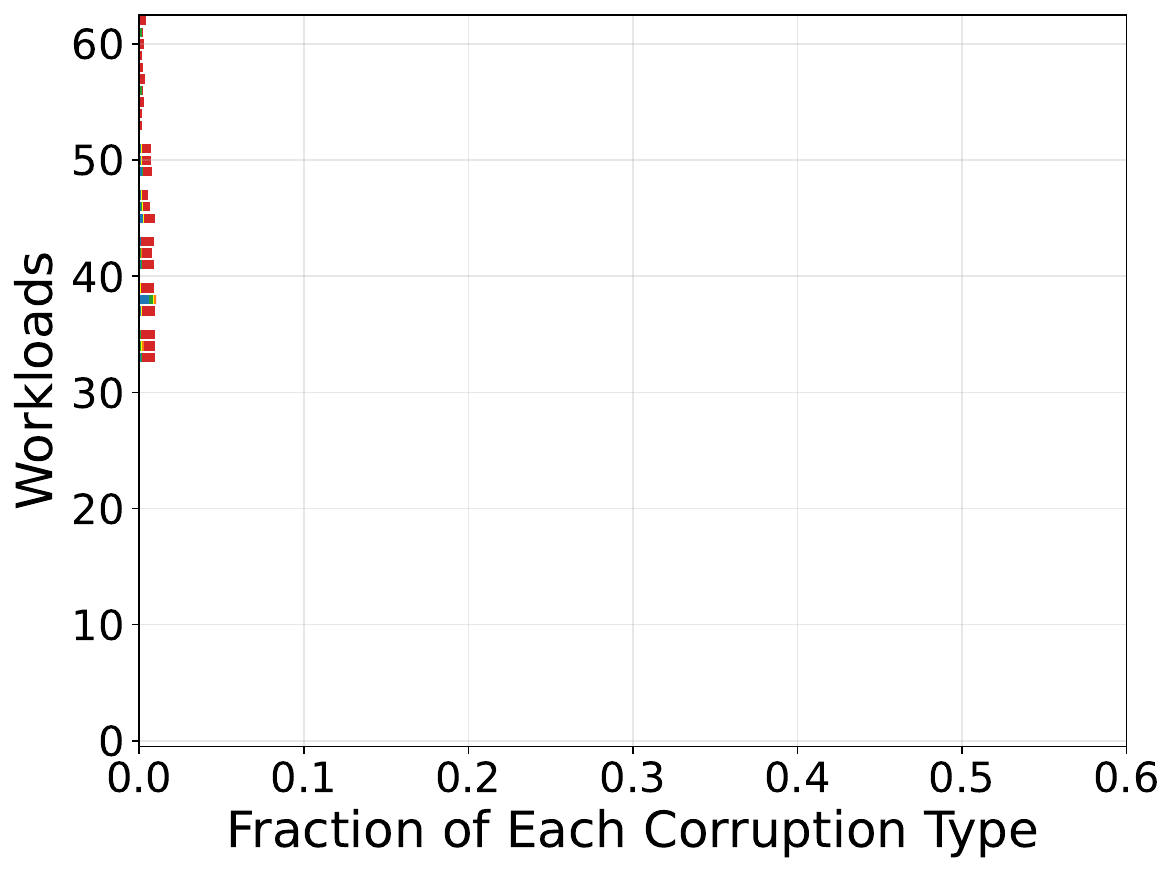}
        \label{subfig: norm_surface_mma}}
    \vspace{-5mm}
    \subfloat[ALU]{
        \includegraphics[width=0.46\columnwidth]{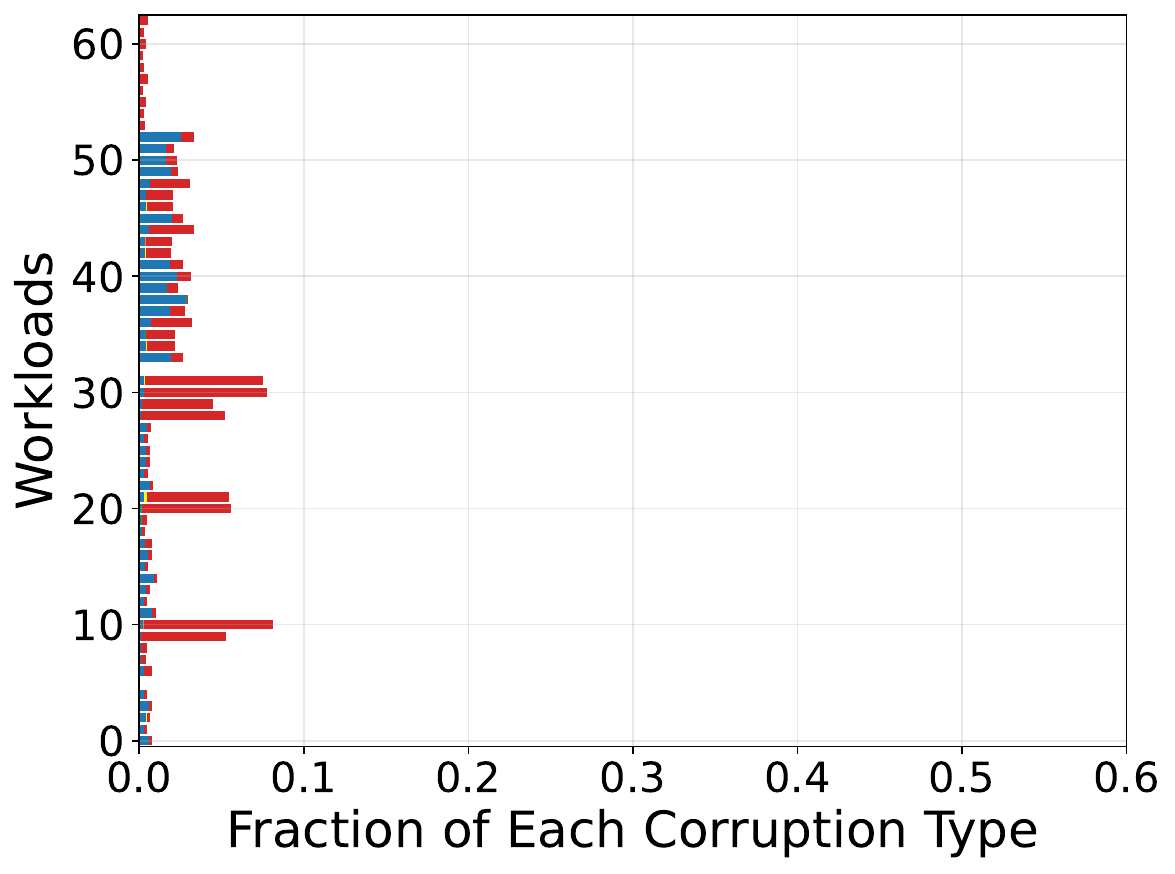}
        \label{subfig: norm_surface_alu}}
    \vspace{2mm}
    \subfloat[L1\$Data]{
        \includegraphics[width=0.46\columnwidth]{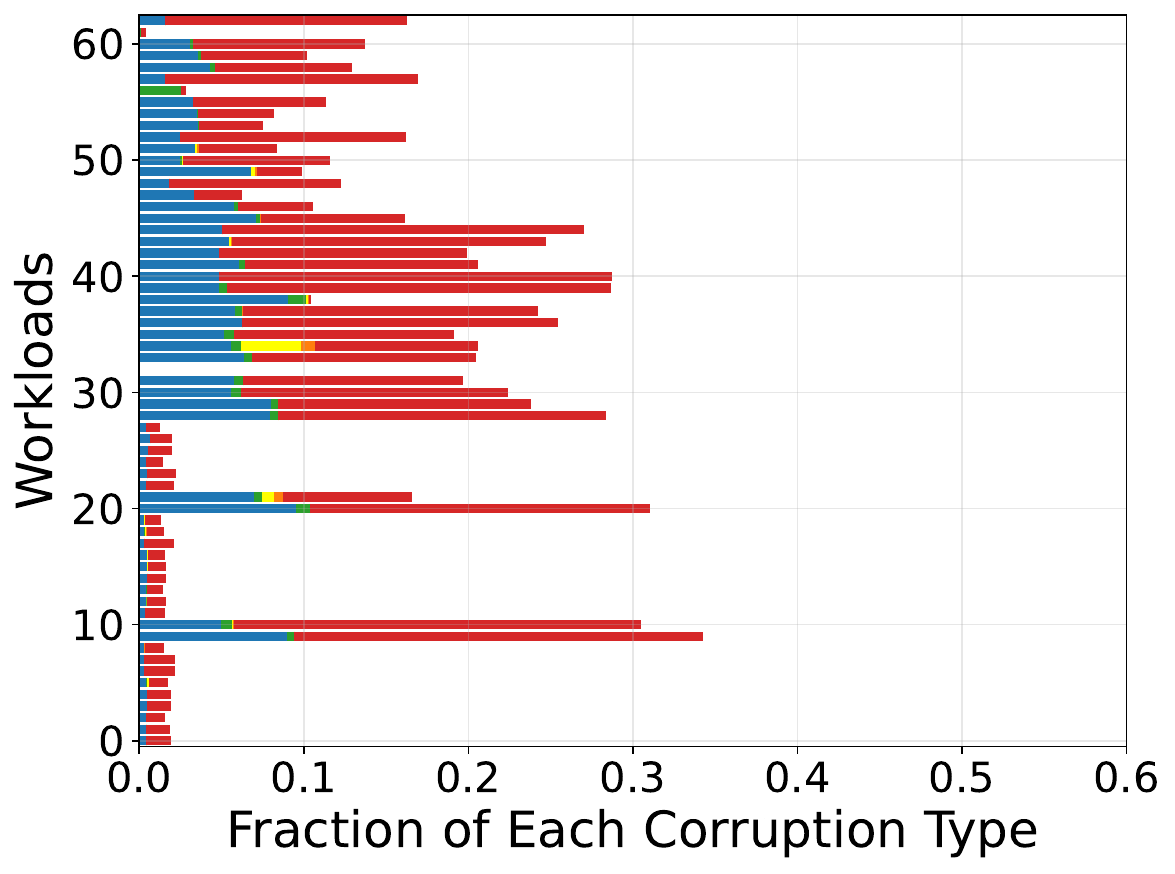}
        \label{subfig: norm_surface_l1data}}
    \vspace{-5mm}
    \subfloat[L1\$Miss Handler]{
        \includegraphics[width=0.46\columnwidth]{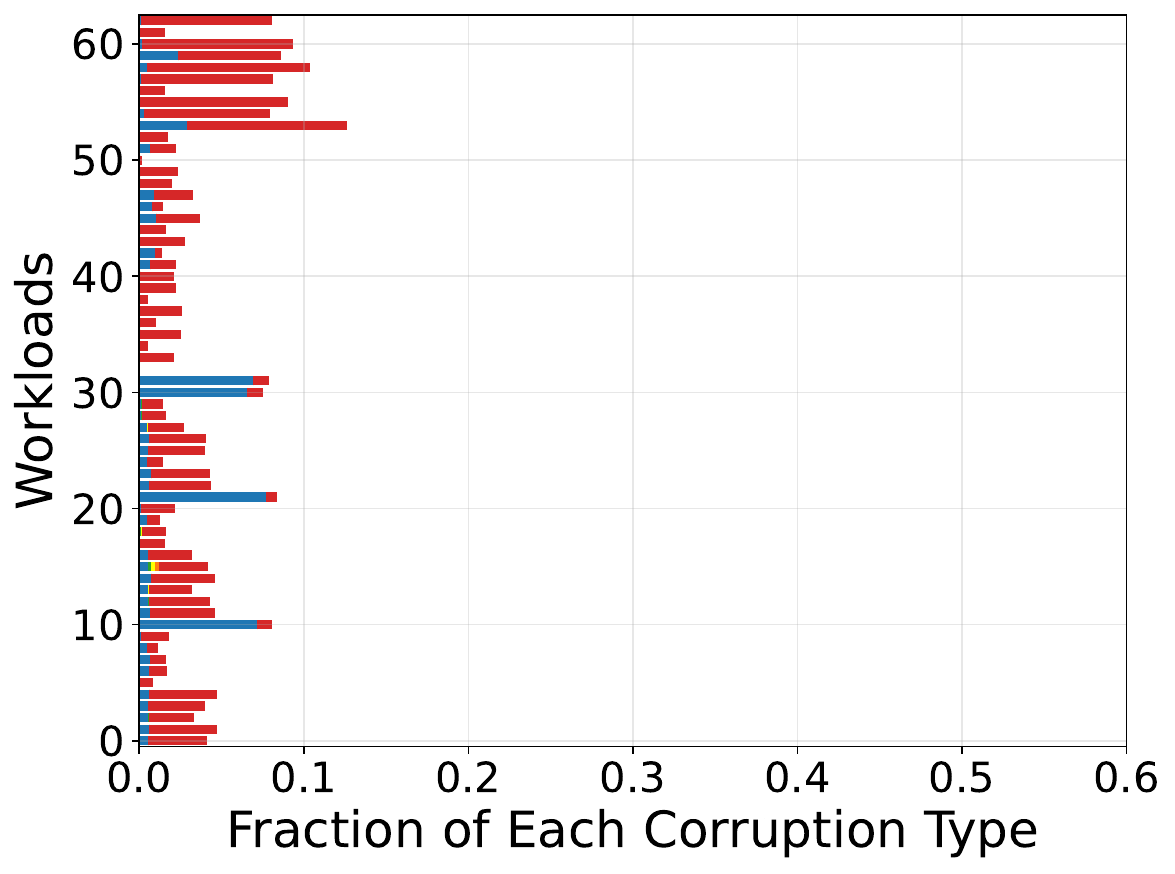}
        \label{subfig: norm_surface_l1miss}}
    \vspace{2mm}
    \subfloat[L1\$Tag]{
        \includegraphics[width=0.46\columnwidth]{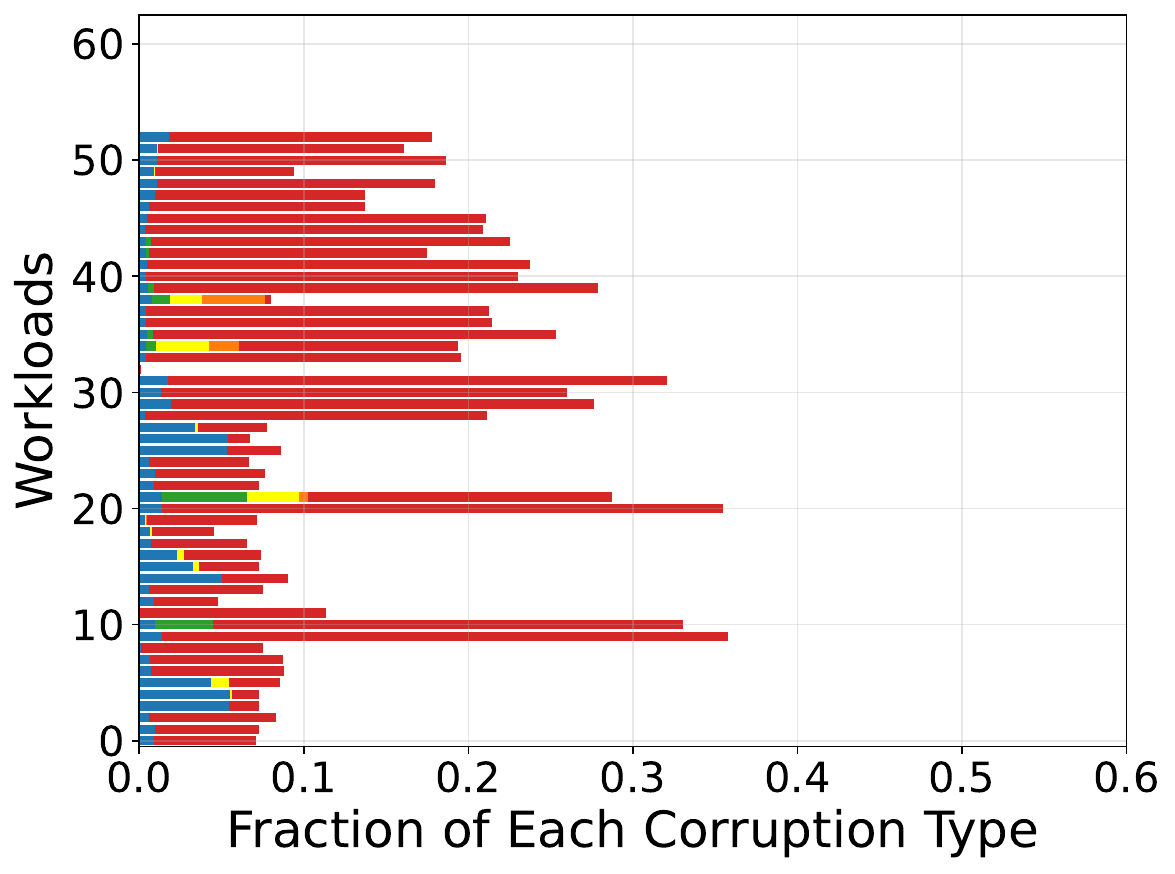}
        \label{subfig: norm_surface_l1tag}}
    \vspace{-2mm}
    \subfloat[CudaCoreIO]{
        \includegraphics[width=0.46\columnwidth]{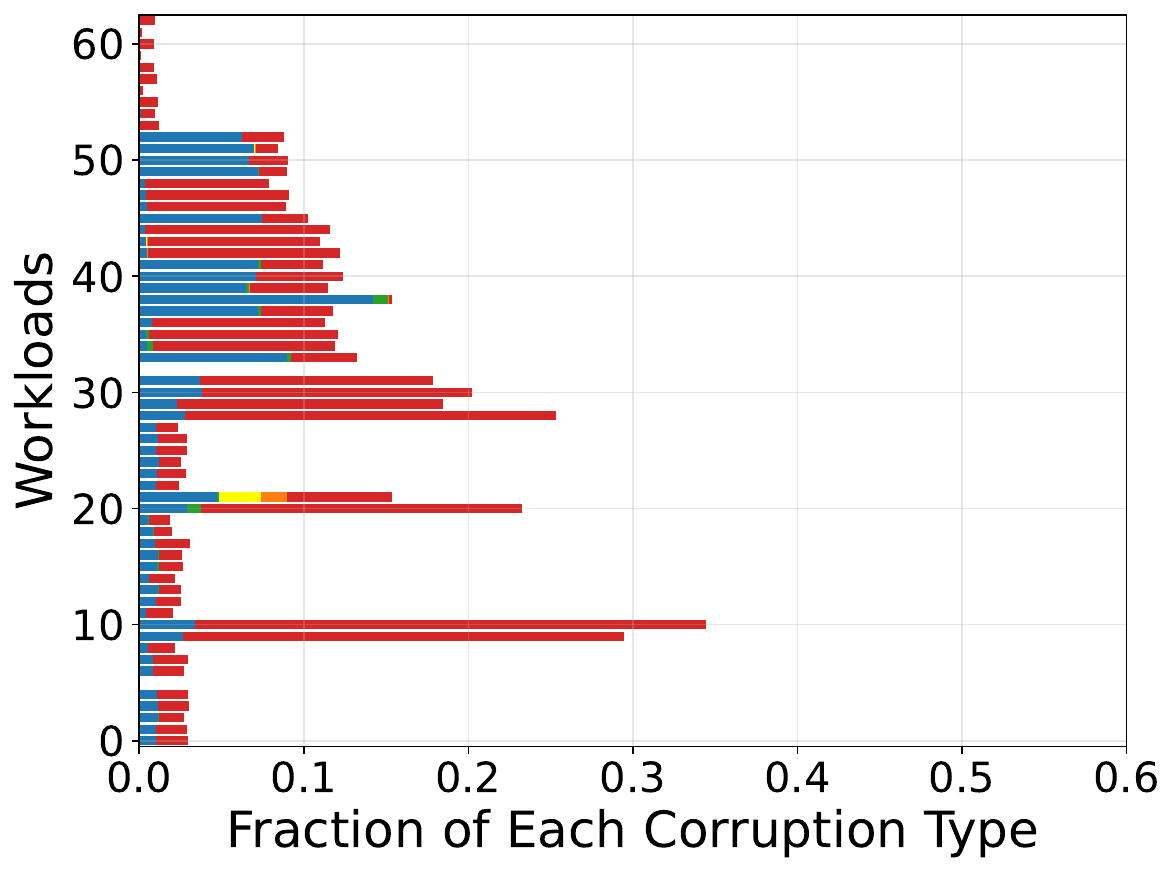}
        \label{subfig: norm_surface_miop}}
    \caption{The corruption type when faults are in specific hardware units. Normalized to the memory space size.}
    \label{fig: corrupt_type_norm_surface}
    \vspace{-3mm}
\end{figure}

Faults are further grouped by their located hardware unit into \emph{CudaCoreControl1}, \emph{CudaCoreControl2}, \emph{TensorCore}, \emph{ALU}, \emph{L1\$Data}, \emph{L1\$Miss Handler}, \emph{L1\$Tag}, and \emph{CudaCoreIO}.
Fig.~\ref{fig: corrupt_type_norm_surface} presents the same data as Fig.~\ref{fig: corrupt_all_units}\subref{subfig: corrupt_type_norm_surface} but split by hardware unit; empty bars indicate no sensitized faults for that benchmark--unit pair.
With a single fault injected in a two-SM netlist, a 50\% error rate often indicates that one SM's output space is entirely corrupted.



%
%

\myobservation{Control and data-buffer faults have high observability; compute faults are subtler.}

\begin{figure}[!t]
    \centering
    \includegraphics[width=0.98\columnwidth]{figs/paper_corruption_type_legend.pdf}
    \vspace{-3mm}

    \subfloat[CudaCoreControl1]{
        \includegraphics[width=0.46\columnwidth]{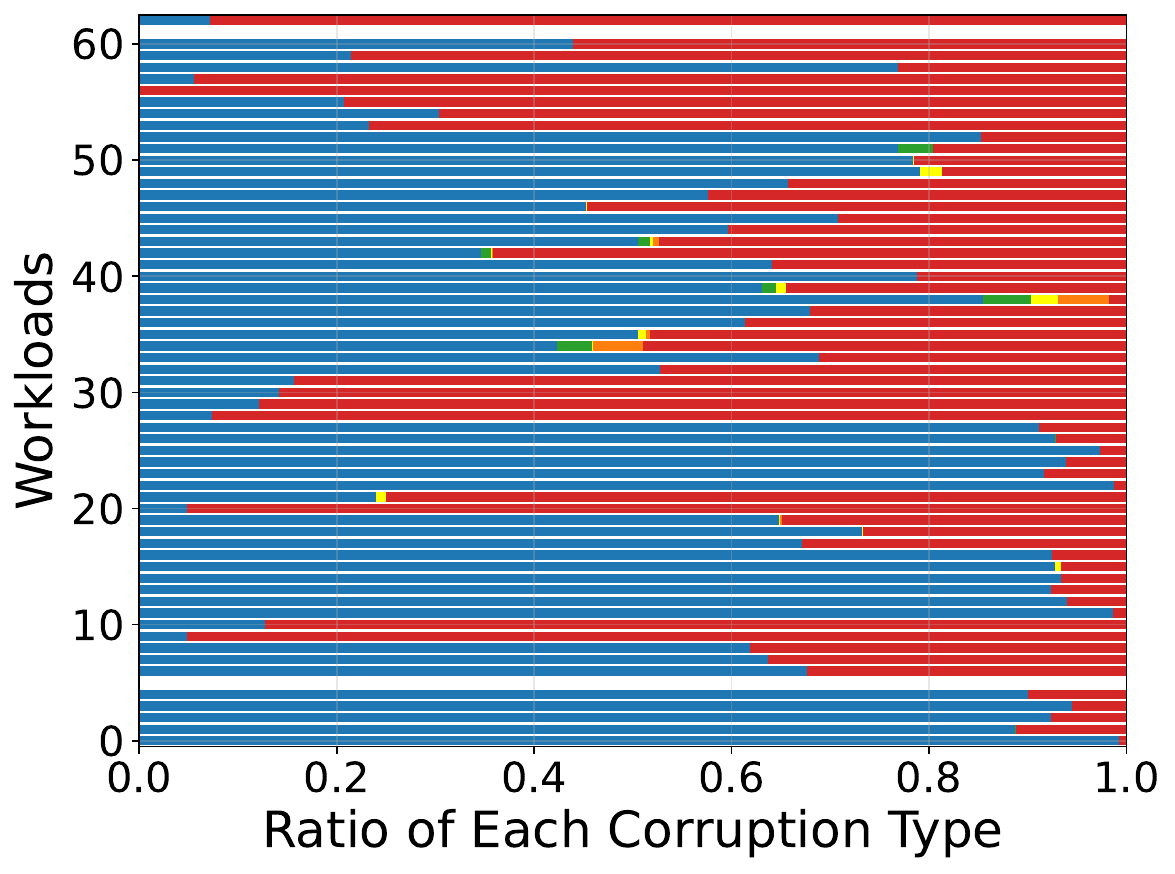}
        \label{subfig: norm_1_sctl}}
    \vspace{-5mm}
    \subfloat[CudaCoreControl2]{
        \includegraphics[width=0.46\columnwidth]{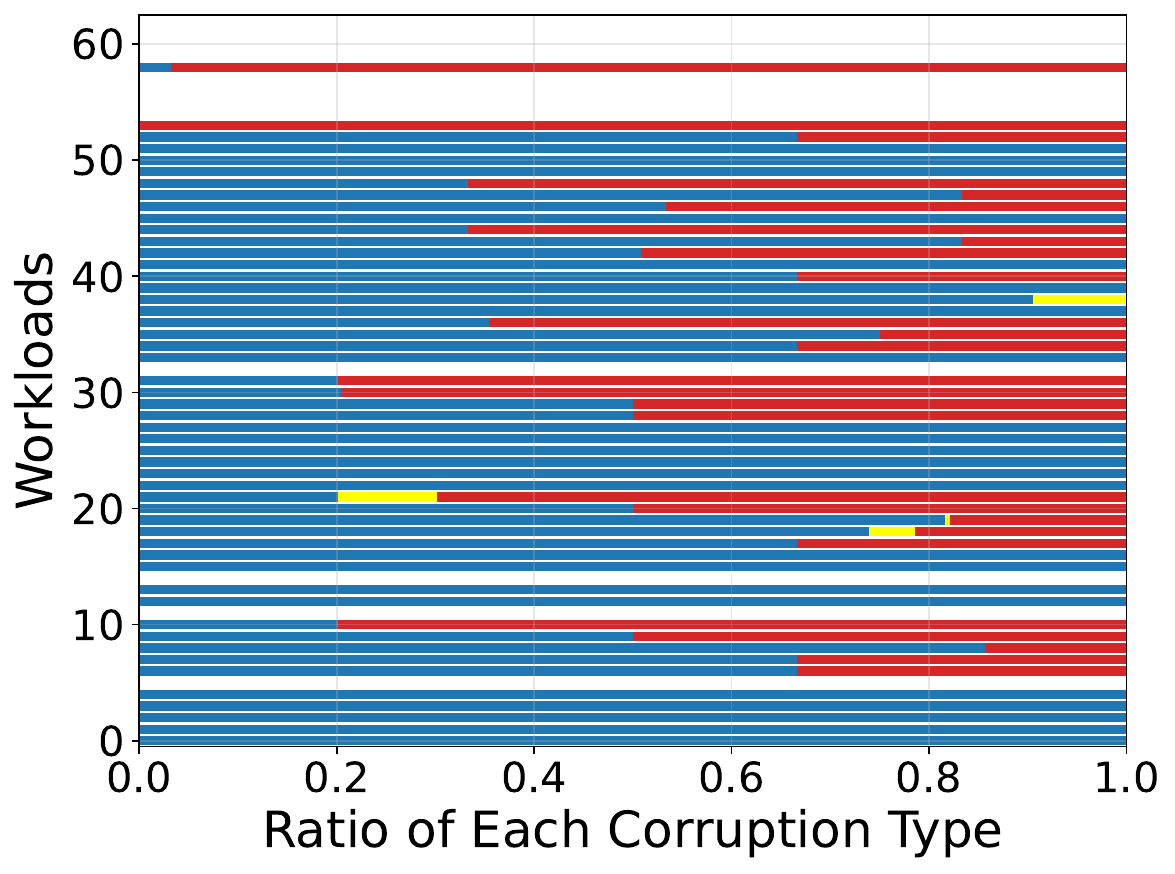}
        \label{subfig: norm_1_sfe}}
    \vspace{2mm}
    \subfloat[TensorCore]{
        \includegraphics[width=0.46\columnwidth]{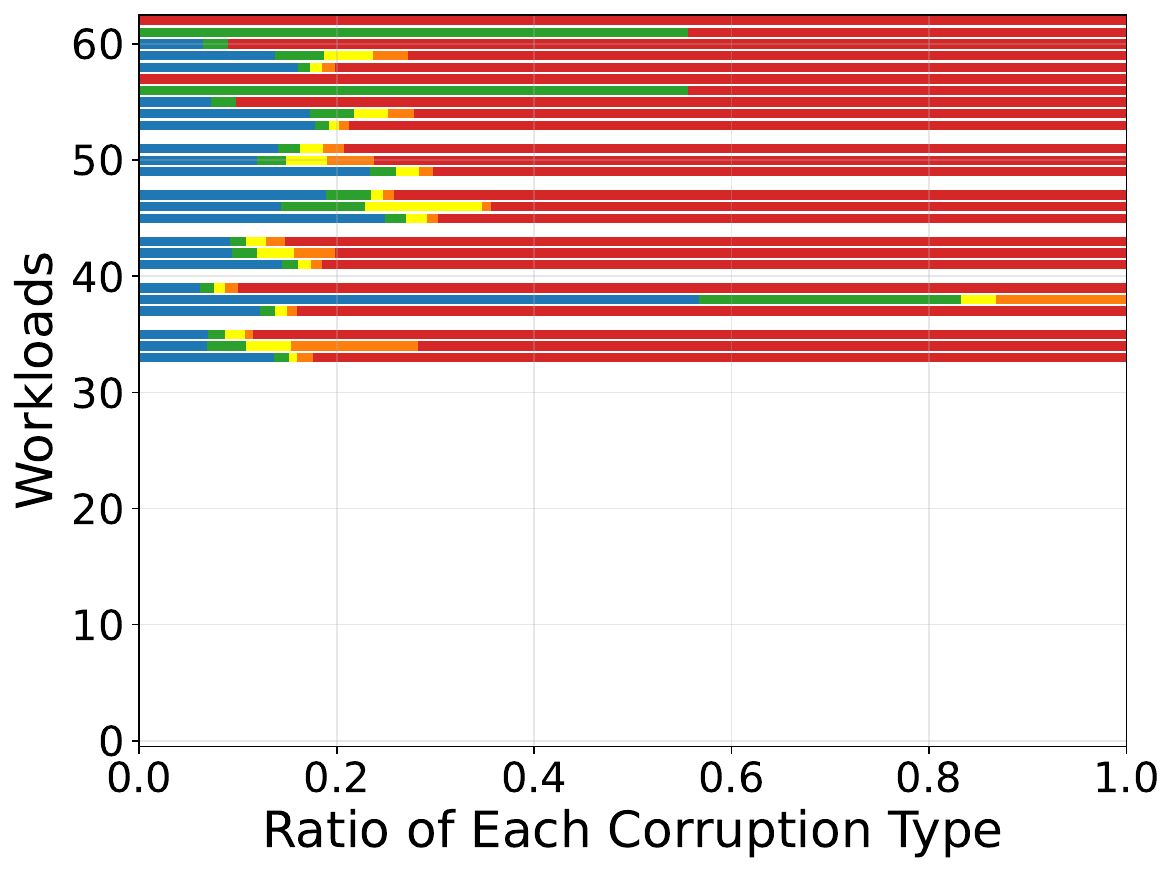}
        \label{subfig: norm_1_mma}}
    \vspace{-5mm}
    \subfloat[ALU]{
        \includegraphics[width=0.46\columnwidth]{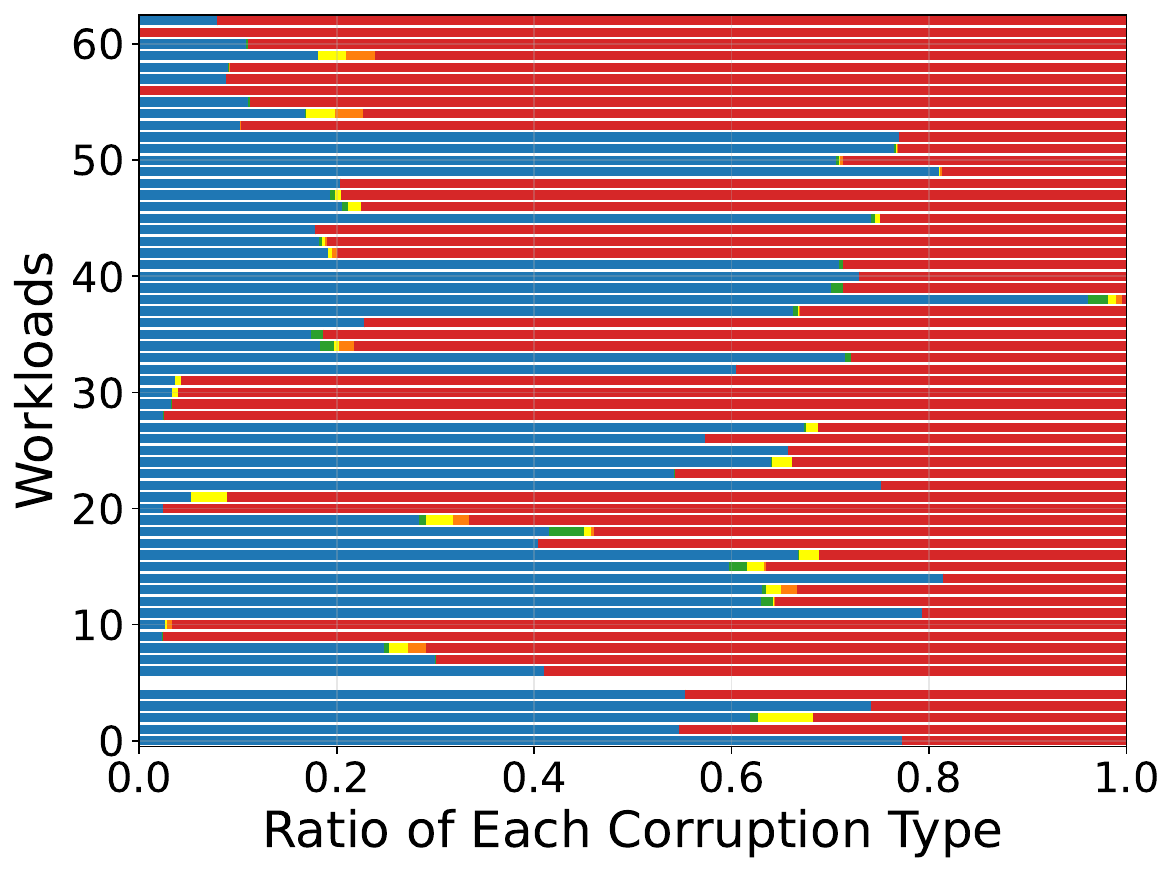}
        \label{subfig: norm_1_alu}}
    \vspace{2mm}
    \subfloat[L1\$Data]{
        \includegraphics[width=0.46\columnwidth]{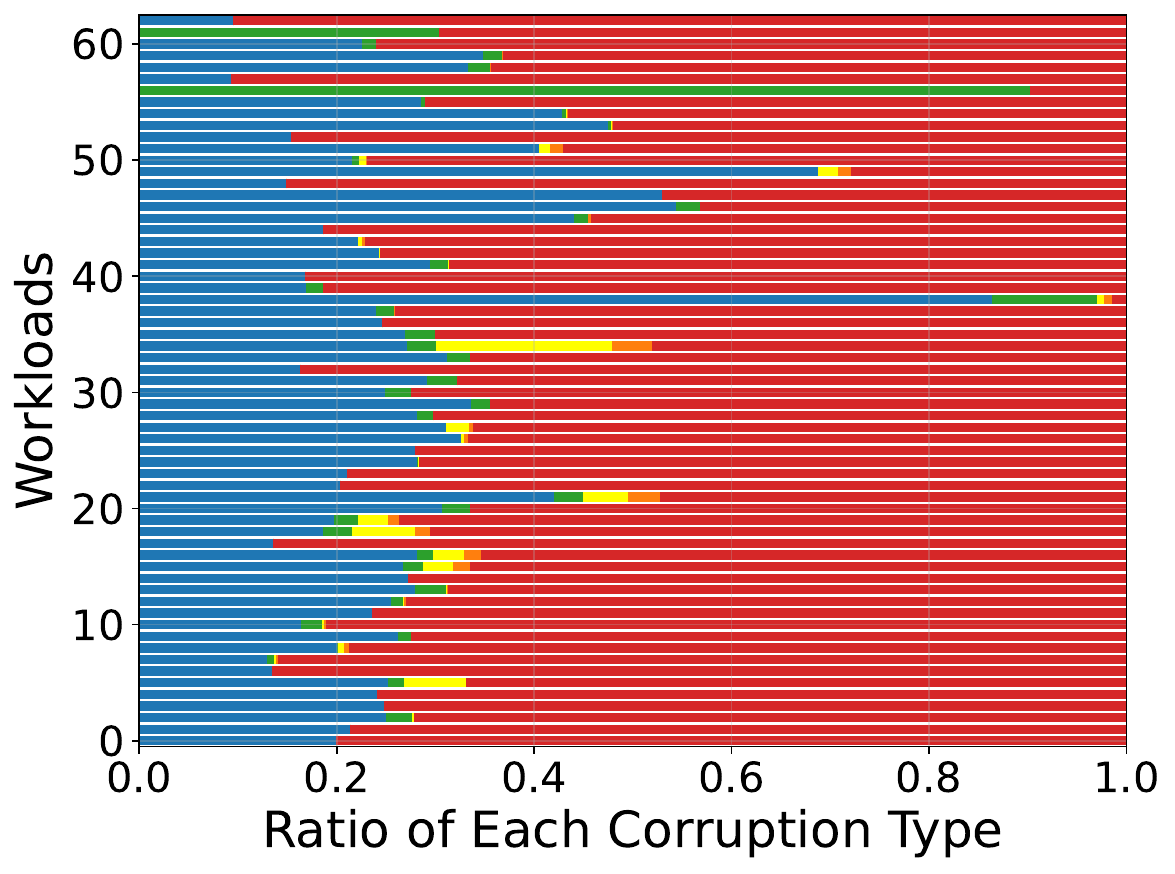}
        \label{subfig: norm_1_l1data}}
    \vspace{-5mm}
    \subfloat[L1\$Miss Handler]{
        \includegraphics[width=0.46\columnwidth]{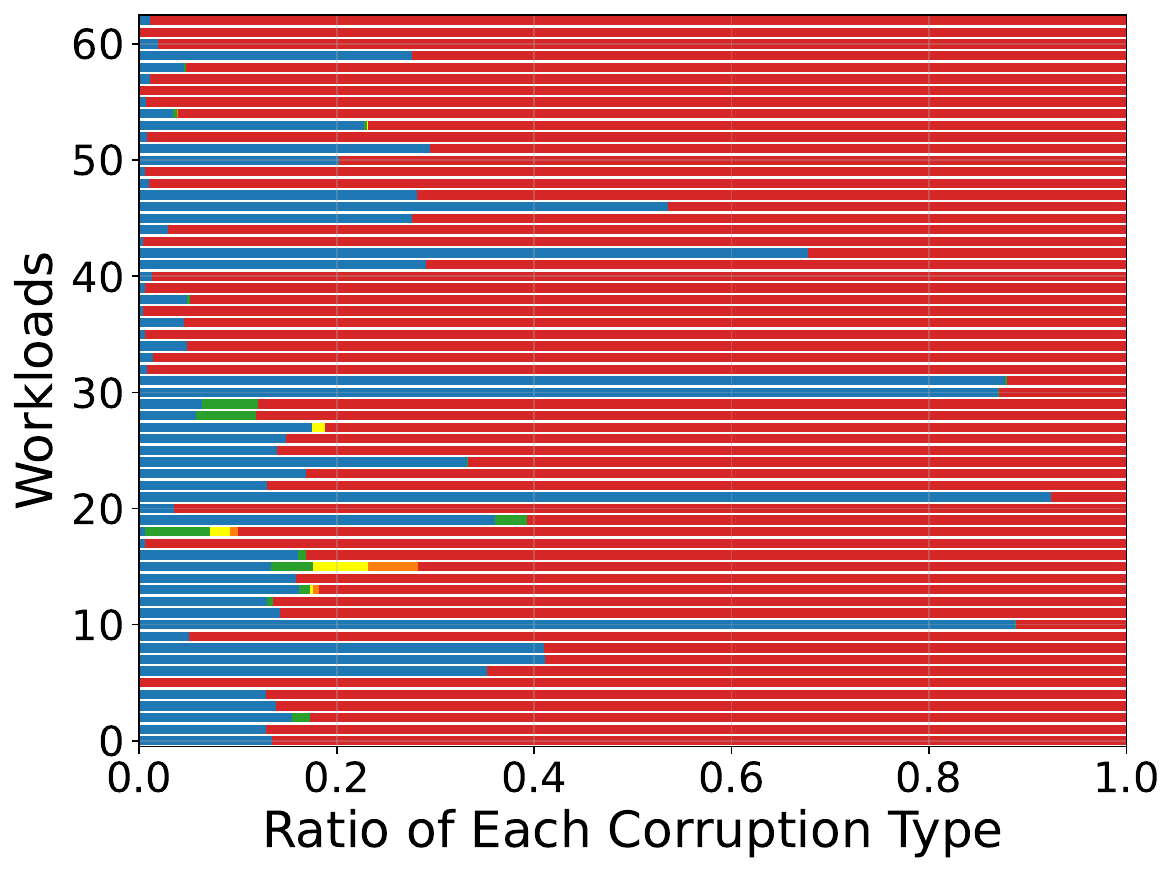}
        \label{subfig: norm_1_l1miss}}
    \vspace{2mm}
    \subfloat[L1\$Tag]{
        \includegraphics[width=0.46\columnwidth]{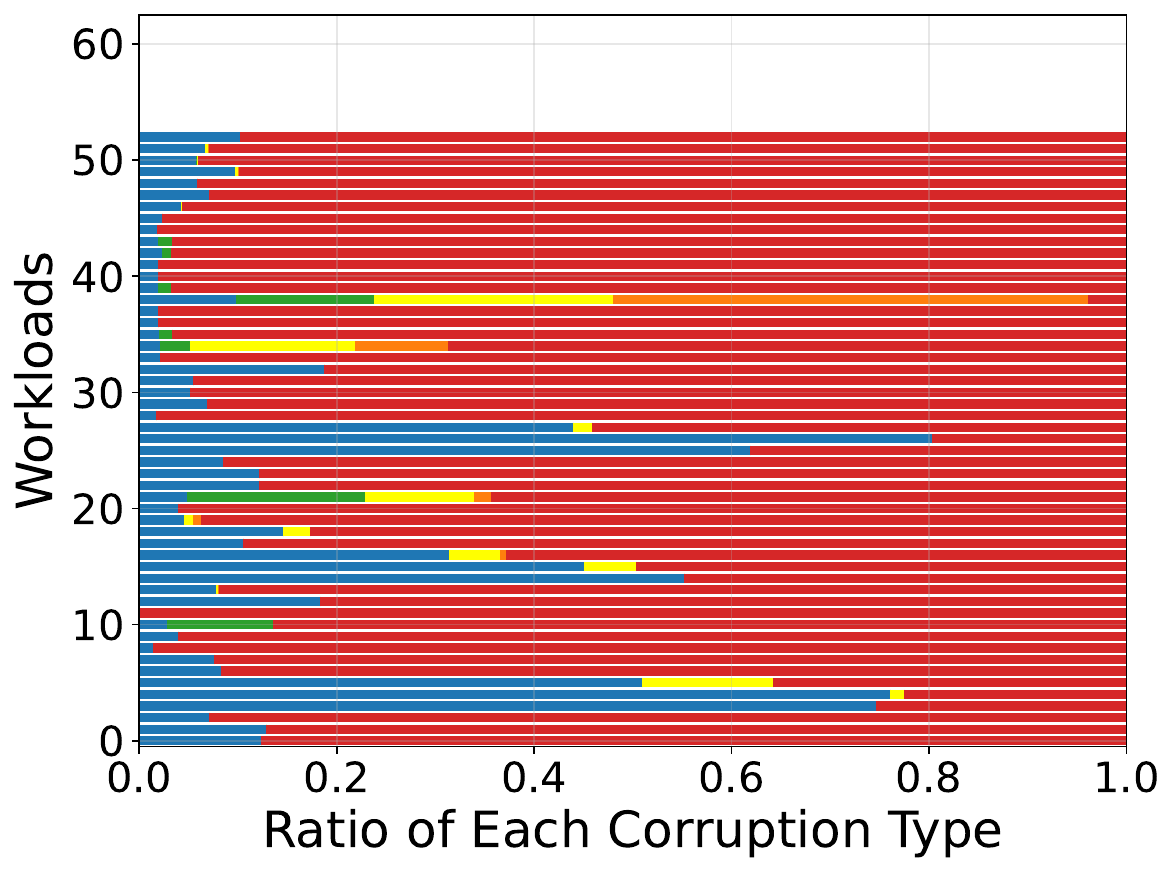}
        \label{subfig: norm_1_l1tag}}
    \vspace{-2mm}
    \subfloat[CudaCoreIO]{
        \includegraphics[width=0.46\columnwidth]{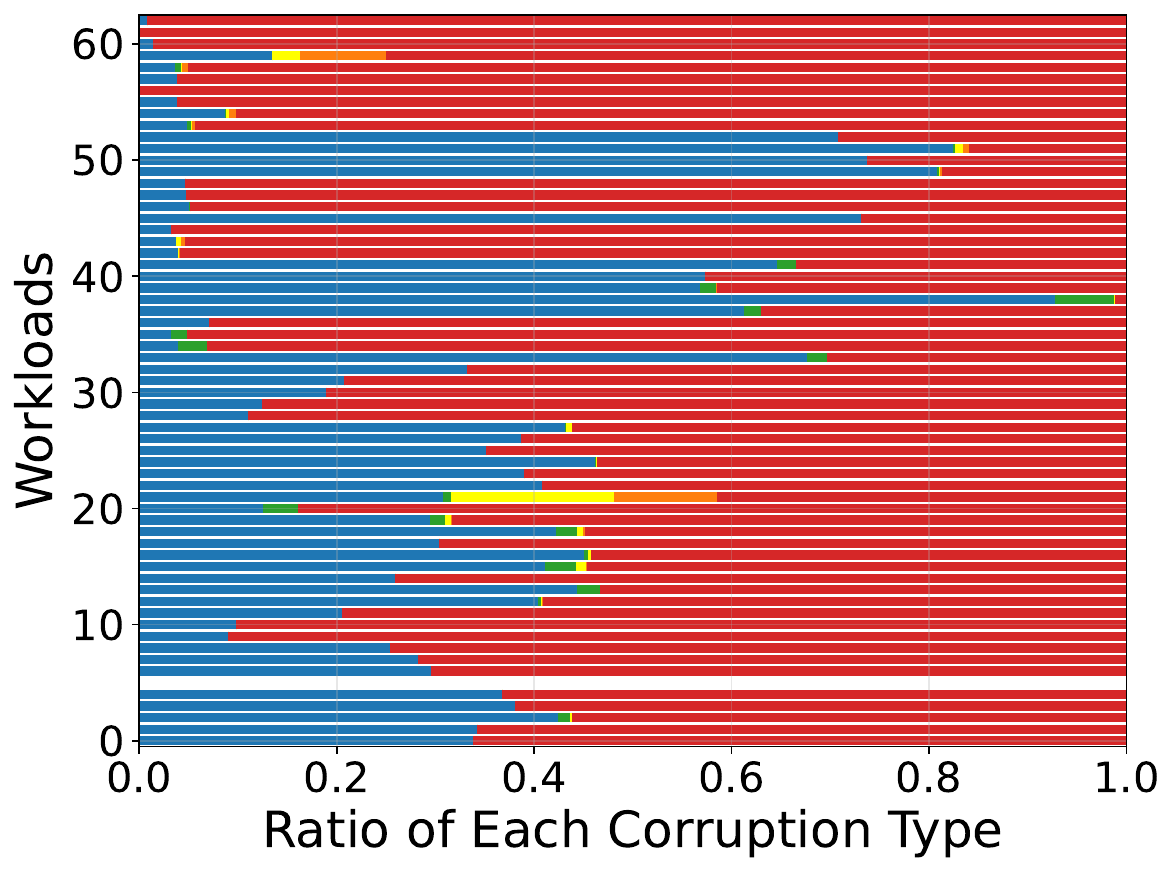}
        \label{subfig: norm_1_miop}}
    \caption{The corruption type when faults are in specific hardware units. Normalized to error rate = 1.}
    \label{fig: corrupt_type_norm_1}
    \vspace{-5mm}
\end{figure}

Although faults in the control logic are often caught as DUE, any escape can lead to a catastrophic impact~\cite{he2020fidelity}.
Fig.~\ref{fig: corrupt_type_norm_surface}\subref{subfig: norm_surface_sctl}\subref{subfig: norm_surface_sfe} show that \emph{CudaCoreControl1} and \emph{CudaCoreControl2} faults often corrupt 20--75\% of an SM's output.
This is because the control logic is responsible for GPU task scheduling, and a fault amplifies its impact by corrupting multiple tasks.
Similarly, data-buffer faults, i.e., in \emph{L1\$Data, L1\$Miss Handler, L1\$Tag, and CudaCoreIO} (Fig.~\ref{fig: corrupt_type_norm_surface}\subref{subfig: norm_surface_l1data}-\subref{subfig: norm_surface_miop}), show higher SDC rates because they contaminate many values in flight.
Conversely, compute-unit faults (\emph{TensorCore} and \emph{ALU}, Fig.~\ref{fig: corrupt_type_norm_surface}\subref{subfig: norm_surface_mma}\subref{subfig: norm_surface_alu}) typically yield lower per-fault corruption rates.
Note that compute units dominate area and thus fault counts, so their impact remains non-negligible.
Those faults, particularly in \emph{TensorCore}, are more likely to manifest special-value corruptions, as shown in Fig.~\ref{fig: corrupt_type_norm_1}\subref{subfig: norm_1_mma}; Fig.~\ref{fig: corrupt_type_norm_1} shows the ratio of each corruption type across hardware units.

\myobservation{Control faults skew toward nullification; data-buffer faults skew toward valid bit flips.}

Fig.~\ref{fig: corrupt_type_norm_1} shows that control-path faults (\emph{CudaCoreControl1/2}) cause mostly nullification, while compute and buffering units produce more non-special bit flips.
This is because compute and storing errors often alter one of the paralleled data paths, but control faults can corrupt or void the entire operation, keeping the memory space in its initialized zeros.

\subsection{Bit-Level Corruption Structure}\label{subsec: bitflip_stats}


%
%
\begin{figure}[!t]
    \centering
    \includegraphics[width=0.98\columnwidth]{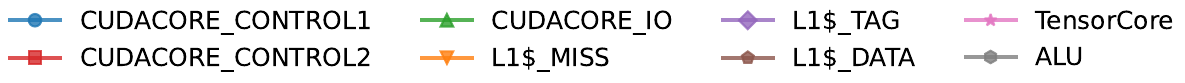}
    \vspace{-3mm}
    
    \subfloat{
        \includegraphics[width=0.98\columnwidth]{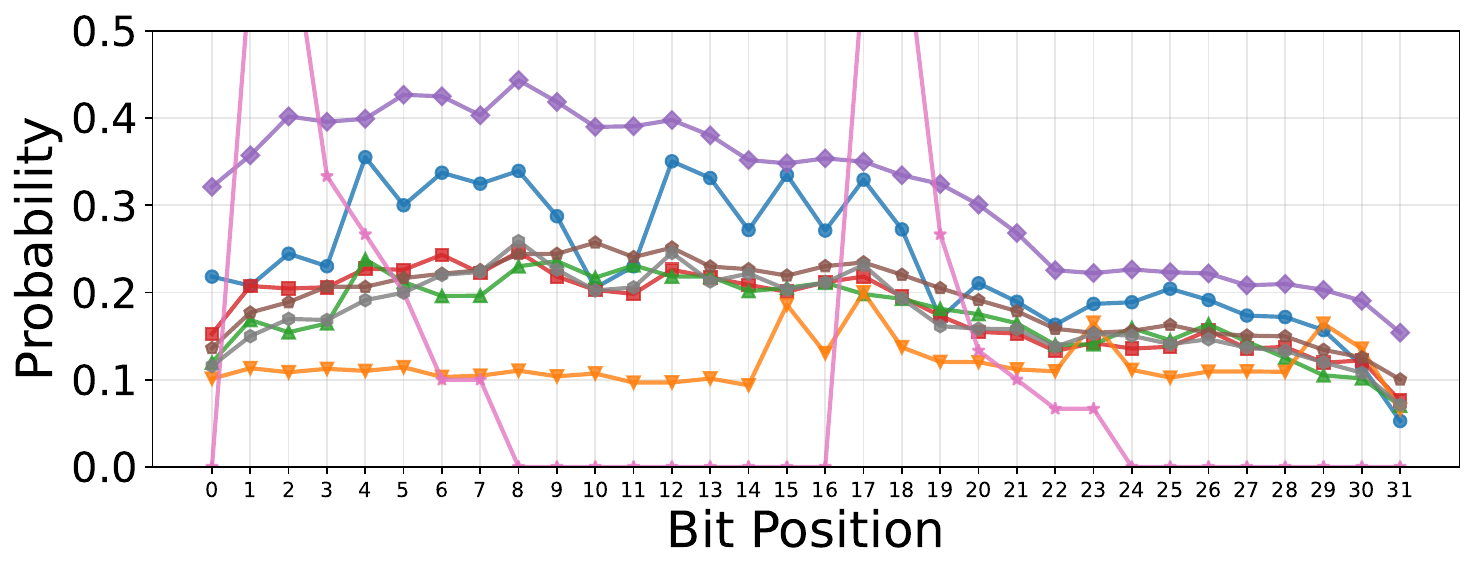}
        \label{subfig: uint32_pos_prob}}
    \vspace{-5mm}
    \subfloat{
        \includegraphics[width=0.98\columnwidth]{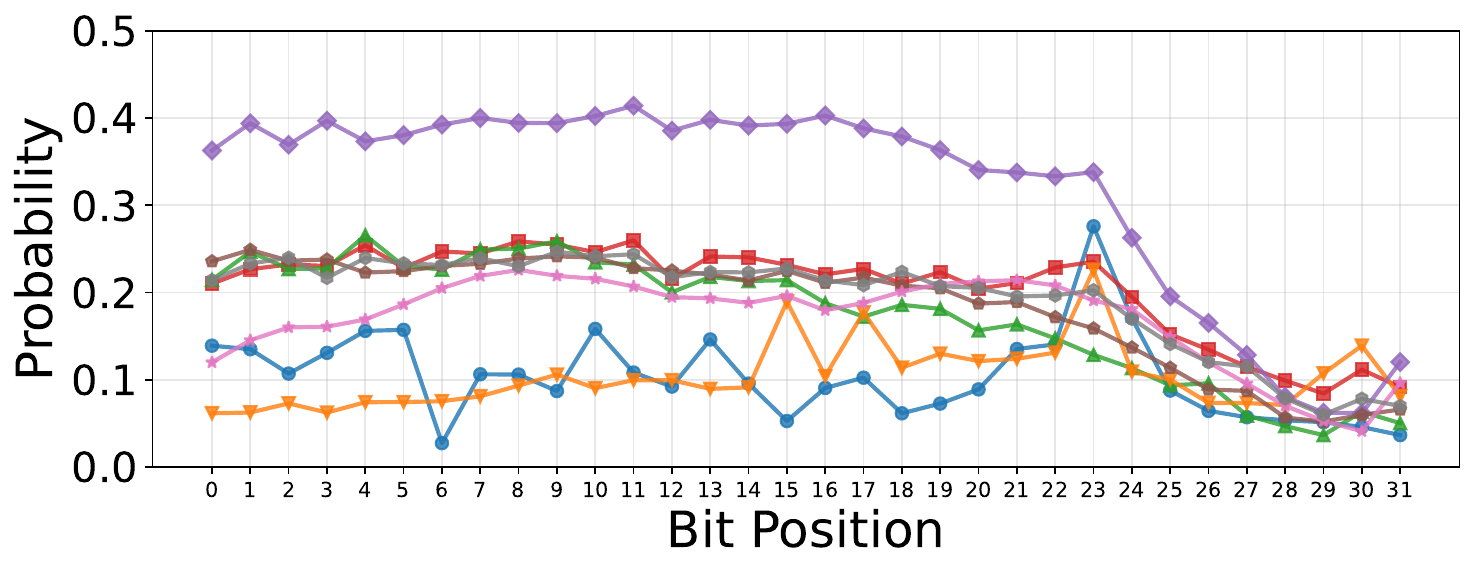}
        \label{subfig: fp32_pos_prob}}
    \vspace{-5mm}
    \subfloat{
        \includegraphics[width=0.98\columnwidth]{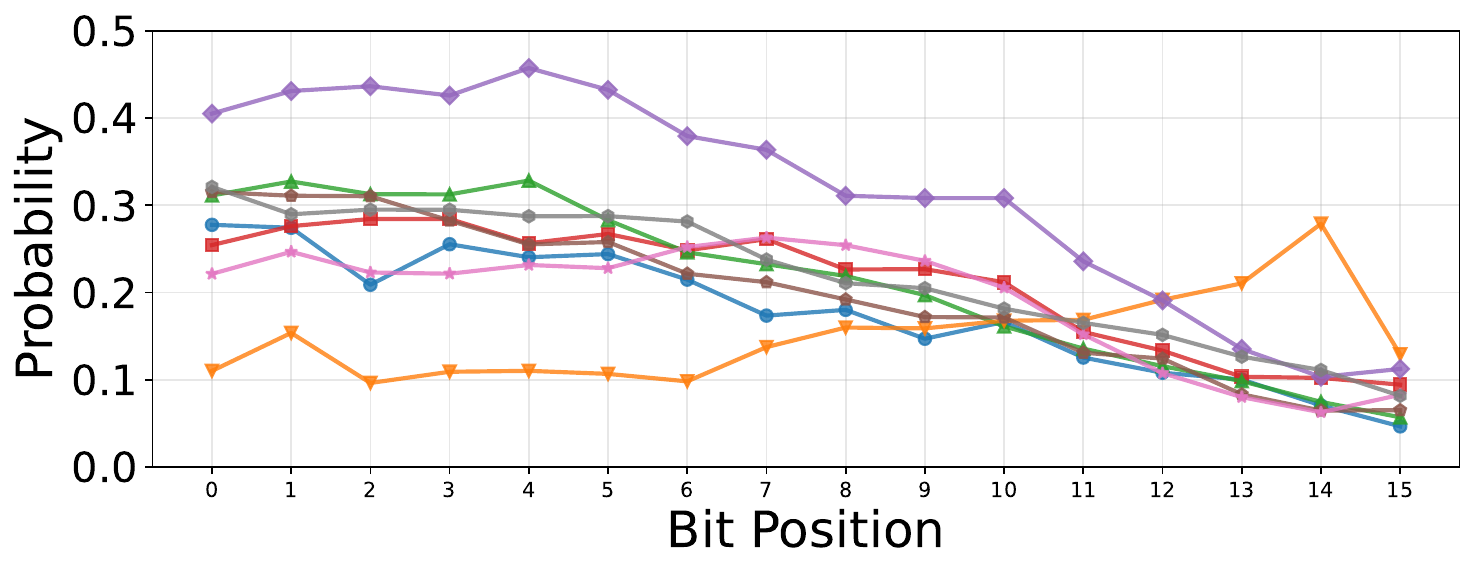}
        \label{subfig: fp16_pos_prob}}
    \caption{The probability of bit-flips of each bit position. Top to bottom: UINT32/FP32/FP16. Bit 0 is the LSB.}
    \label{fig: bit_pos_prob}
    \vspace{-3mm}
\end{figure}

%
%
\begin{figure}[!t]
    \centering
    \includegraphics[width=0.98\columnwidth]{figs/paper_bit_stats_legend.pdf}
    \vspace{-3mm}
    
    \subfloat{
        \includegraphics[width=0.98\columnwidth]{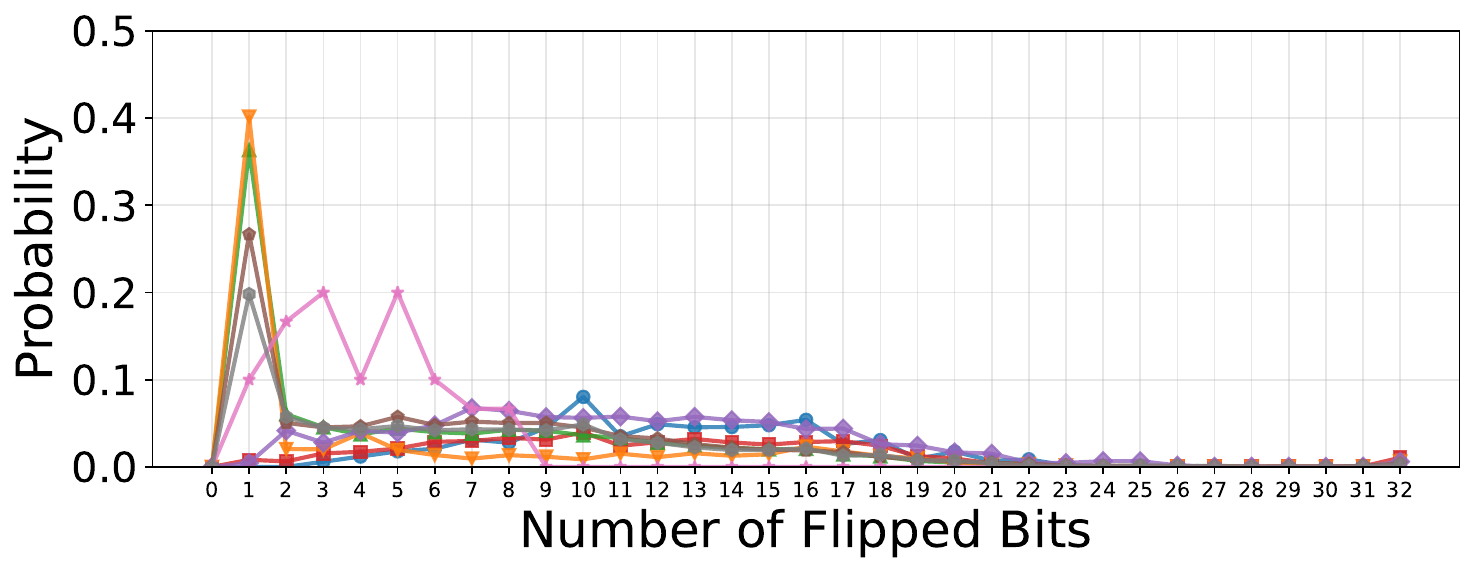}
        \label{subfig: uint32_count_prob}}
    \vspace{-5mm}
    \subfloat{
        \includegraphics[width=0.98\columnwidth]{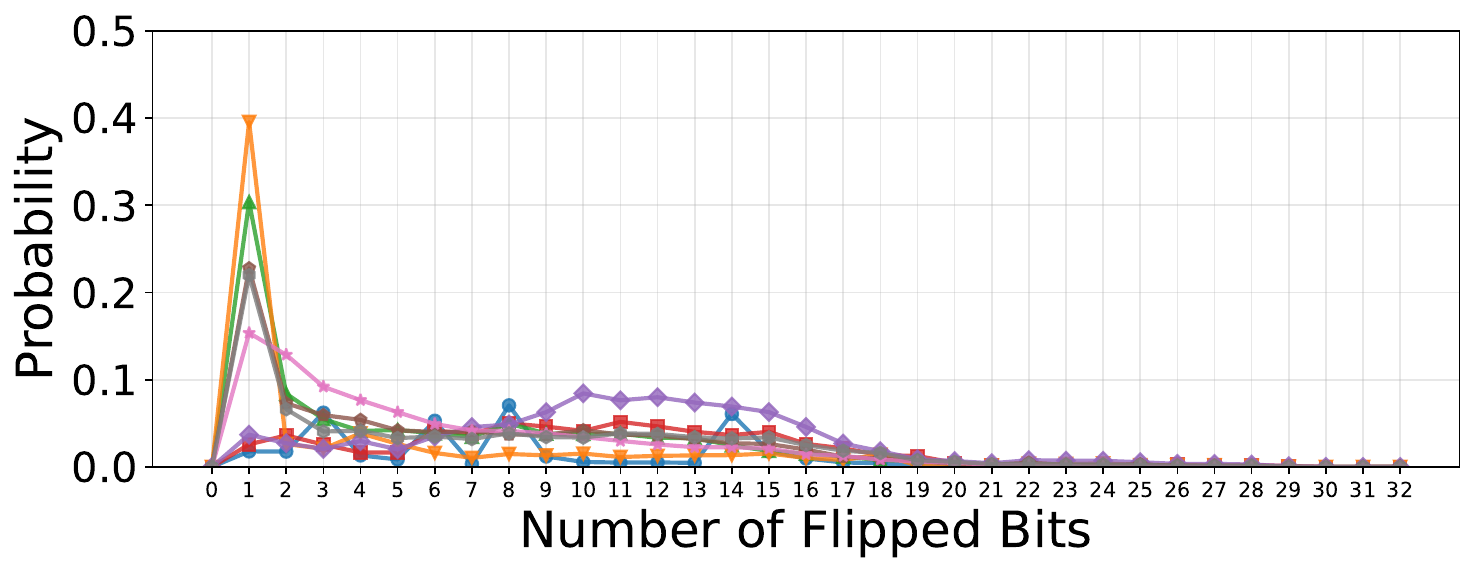}
        \label{subfig: fp32_count_prob}}
    \vspace{-5mm}
    \subfloat{
        \includegraphics[width=0.98\columnwidth]{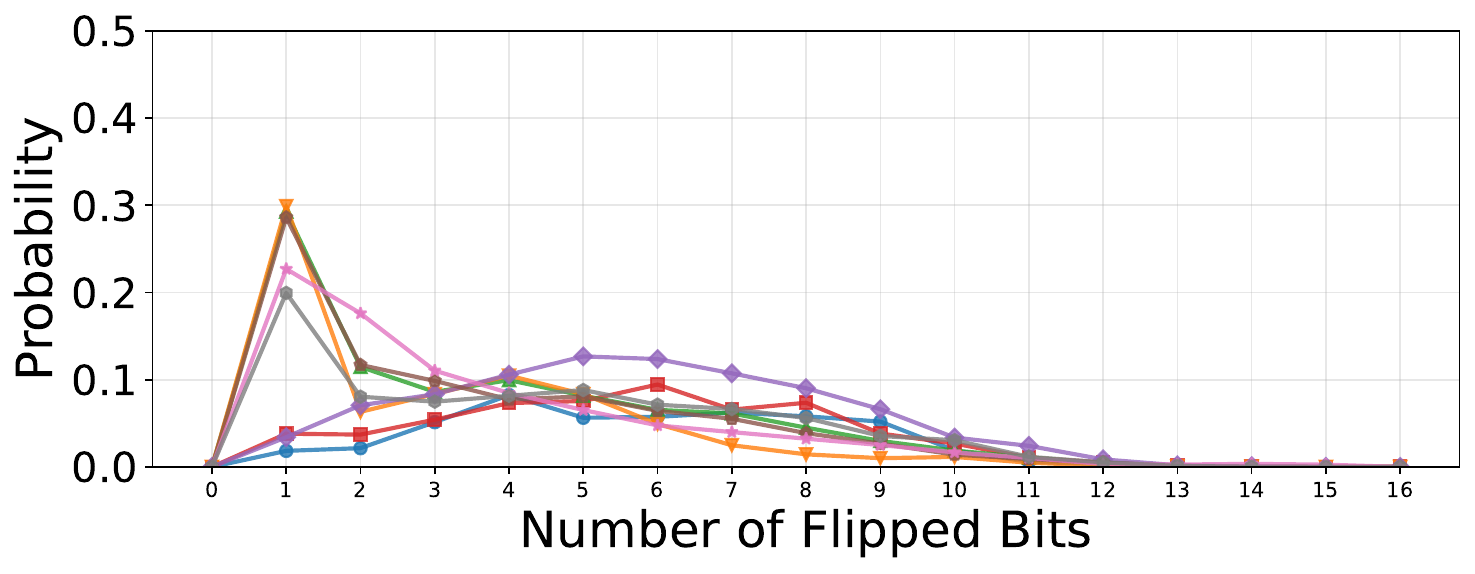}
        \label{subfig: fp16_count_prob}}
    \caption{The probability of bit-flip counts per corrupted value. Top to bottom: UINT32/FP32/FP16.}
    \label{fig: bit_count_prob}
    \vspace{-3mm}
\end{figure}

A majority of SDC manifests as corrupt but valid values (denoted as ``non-special'' in Fig.~\ref{fig: corrupt_all_units}--\ref{fig: corrupt_type_norm_1}), which are general bit-flips besides those with fixed bit patterns, i.e., nullification and NaN/$\pm$INF.
This section presents the aggregated bit-flip statistics in terms of data types and hardware units.
Fig.~\ref{fig: bit_pos_prob} shows the probability of bit-flips of each bit position, and Fig.~\ref{fig: bit_count_prob} shows the number of bit-flips of each corruption, both aggregated across all micro-benchmarks of that output data type.
%
FP8 workloads are included in the overall statistics but omitted from per-format analysis due to limited samples and higher variance; observed trends align with other FP formats.

\myobservation{Bit-flip distribution varies by hardware unit.}


Complementing the overall corruption-type breakdown in Fig.~\ref{fig: corrupt_type_norm_surface}, Fig.~\ref{fig: bit_pos_prob} characterizes bit-flip probabilities conditioned on non-special corruptions.
Higher overall per-position flip rate in Fig.~\ref{fig: bit_pos_prob}, such as \emph{L1\$Tag}, corresponds to increased likelihood of multi-bit corruptions, as reflected in Fig.~\ref{fig: bit_count_prob}.



\myobservation{Bit-flip probability decreases toward MSB.}

Within a corruption, the flip rate across bit positions can be non-uniformly distributed.
In Fig.~\ref{fig: bit_pos_prob}, flip rates drop from LSB to MSB, with FP exponent bits (bits 23--30 for FP32, 10--14 for FP16) least affected, reflecting simpler exponent logic than mantissa data paths.
Lower-precision formats (e.g., BF16 and FP16) are more vulnerable to precision errors due to their reduced mantissa width relative to FP32, while the smaller exponent field in FP16 limits its dynamic range, making it less susceptible to extreme magnitude shifts than FP32 or BF16.

\myobservation{Single-bit flips are not the majority on GPUs.}

Fig.~\ref{fig: bit_count_prob} shows that single-bit peaks are unit-specific and never exceed 40\% of non-special corruptions even when present, contrasting with CPU reports of 72--98\% single-bit flips~\cite{wang2023understanding}.
Multi-bit events instead form a concentrated mode over intermediate flip counts rather than appearing as tail events.

\subsection{Warp-Aligned Spatial Correlation}
\begin{figure*}[t]
  \centering
  \includegraphics[width=0.98\columnwidth]{figs/paper_corruption_type_legend.pdf}
  \vspace{-3mm}

  \setlength{\tabcolsep}{0.1em}
  \begin{tabular}{ccccccccc}
    \vspace{-3mm}
    \raisebox{0.2\height}{\makebox[0pt][r]{\rotatebox{90}{UINT32}}} &
    \subfloat{\includegraphics[width=0.245\columnwidth]{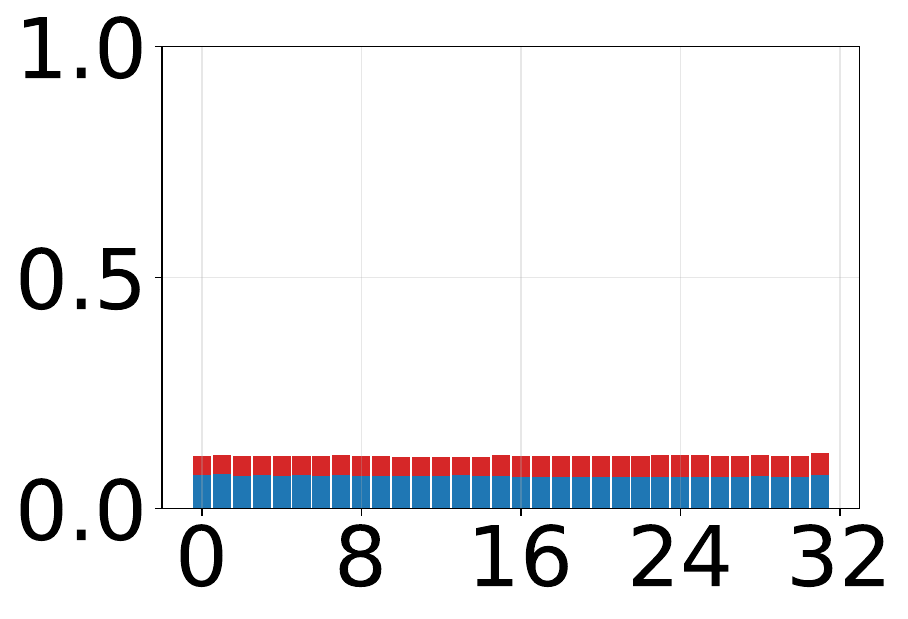}} &
    \subfloat{\includegraphics[width=0.245\columnwidth]{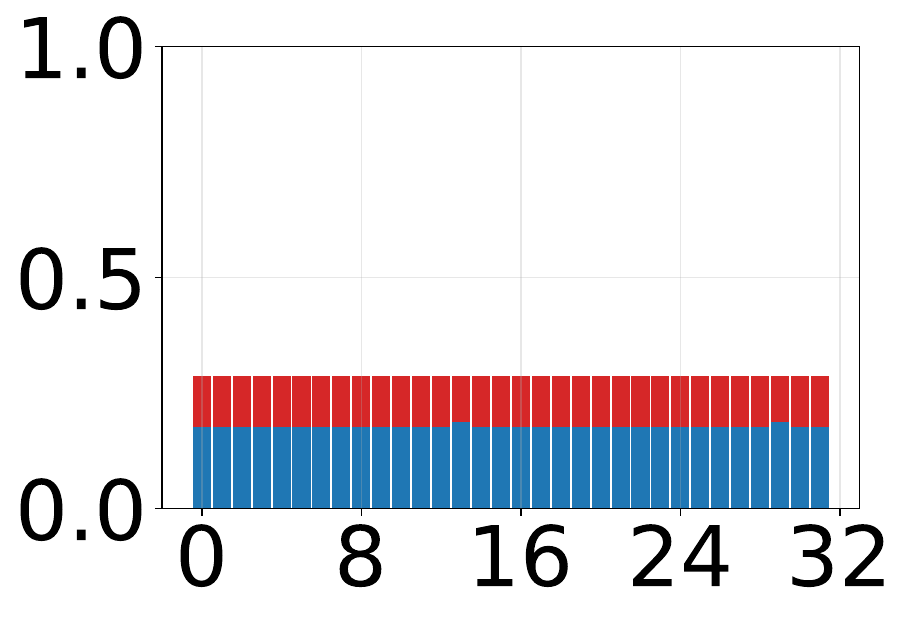}} &
    \subfloat{\includegraphics[width=0.245\columnwidth]{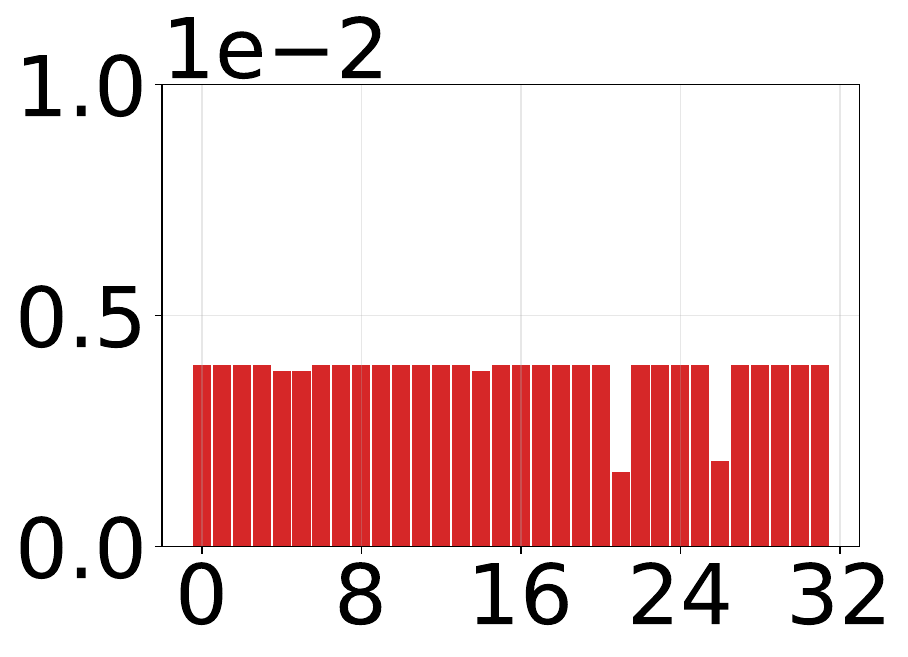}} &
    \subfloat{\includegraphics[width=0.245\columnwidth]{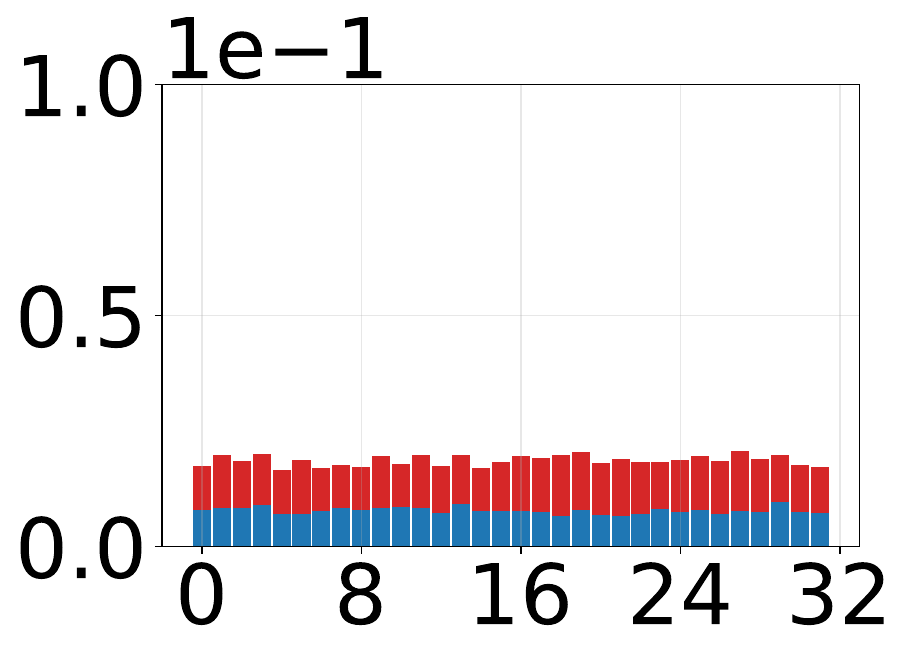}} &
    \subfloat{\includegraphics[width=0.245\columnwidth]{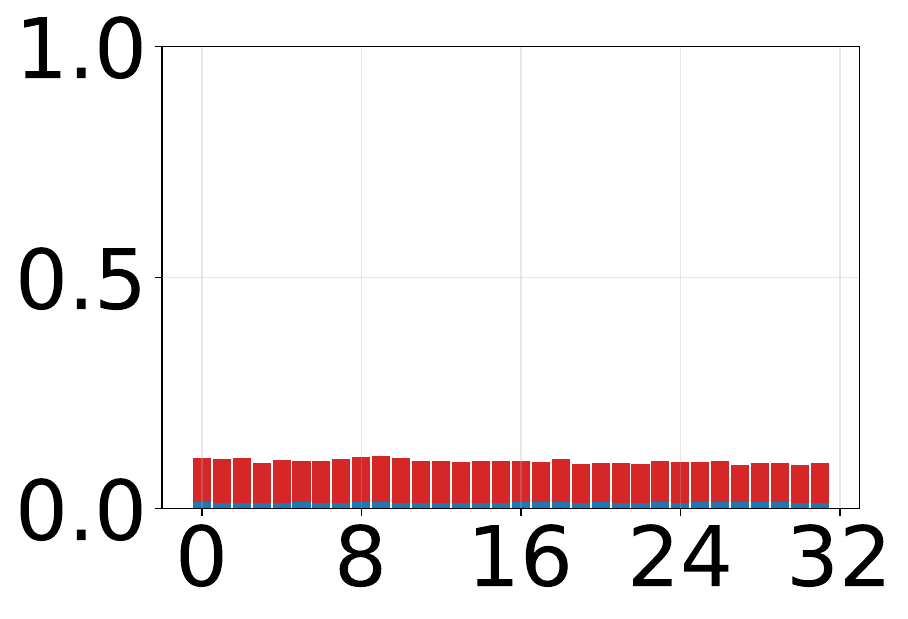}} &
    \subfloat{\includegraphics[width=0.245\columnwidth]{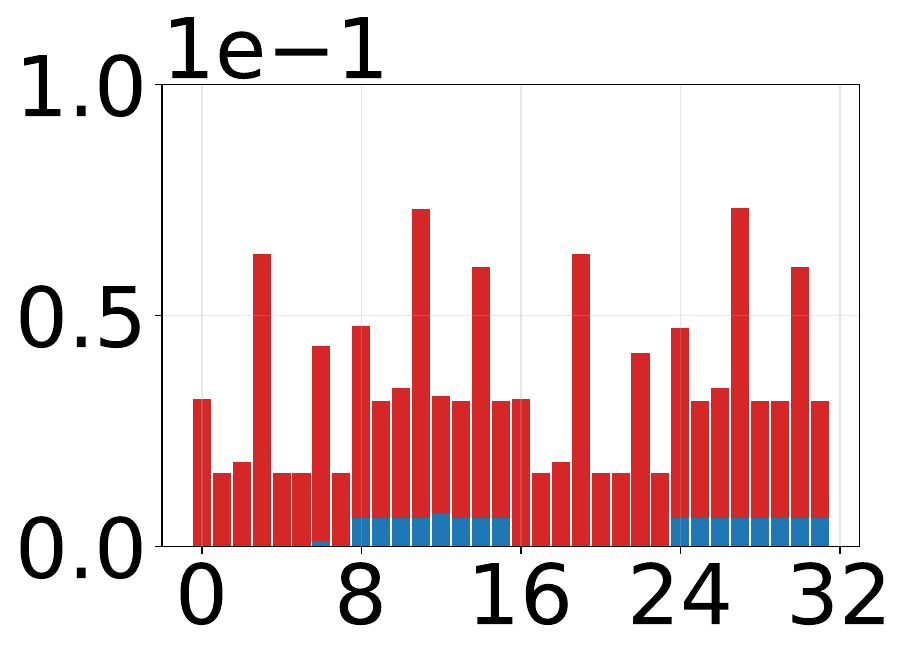}} &
    \subfloat{\includegraphics[width=0.245\columnwidth]{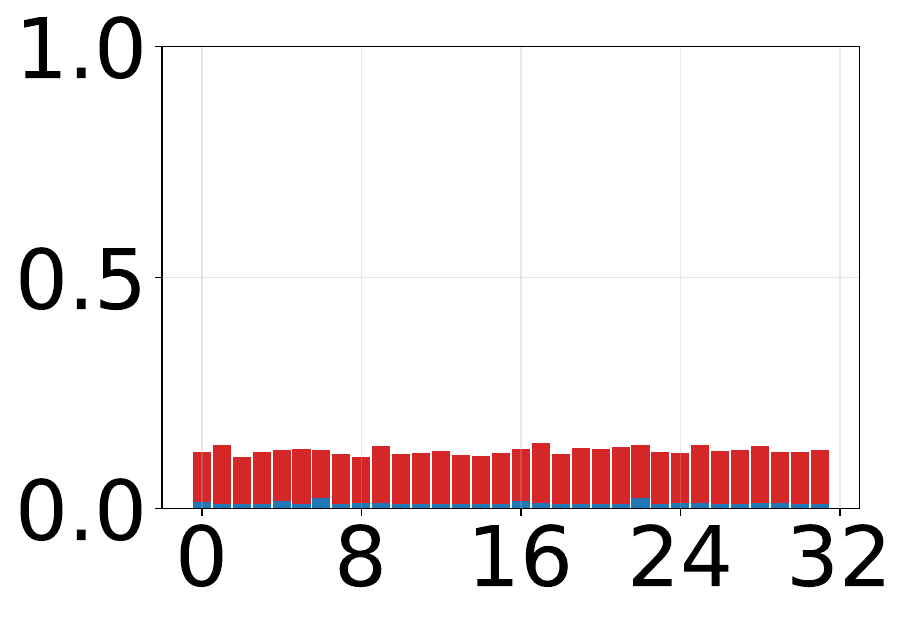}} &
    \subfloat{\includegraphics[width=0.245\columnwidth]{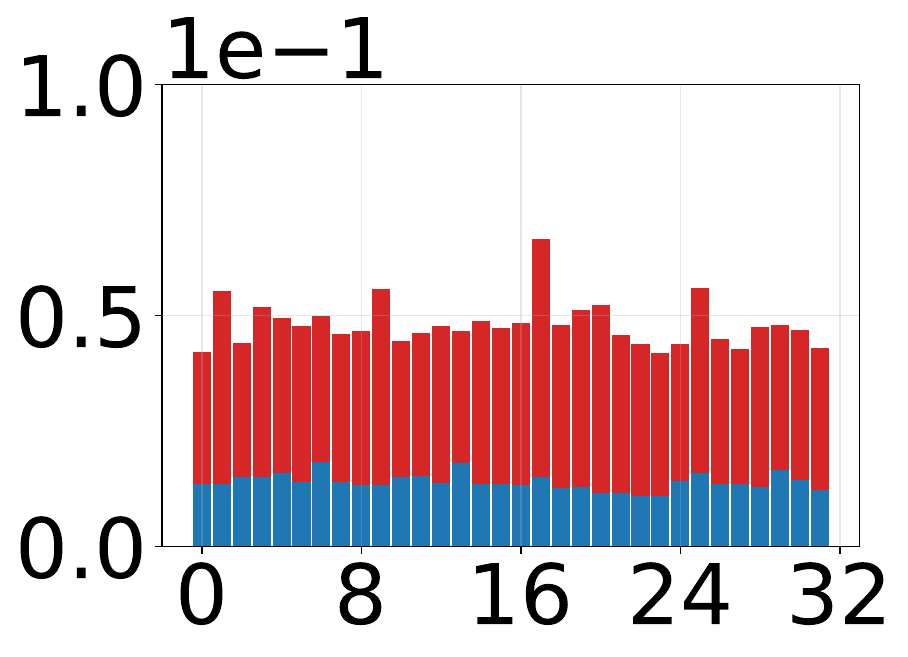}} \\
    \vspace{-3mm}
    
    \raisebox{0.65\height}{\makebox[0pt][r]{\rotatebox{90}{FP32}}} &
    \subfloat{\includegraphics[width=0.245\columnwidth]{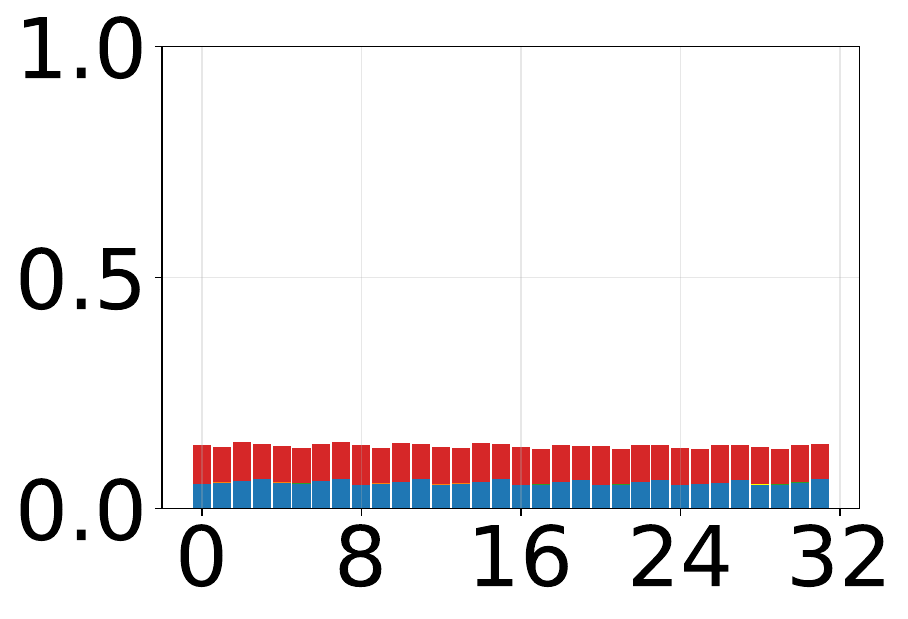}} &
    \subfloat{\includegraphics[width=0.245\columnwidth]{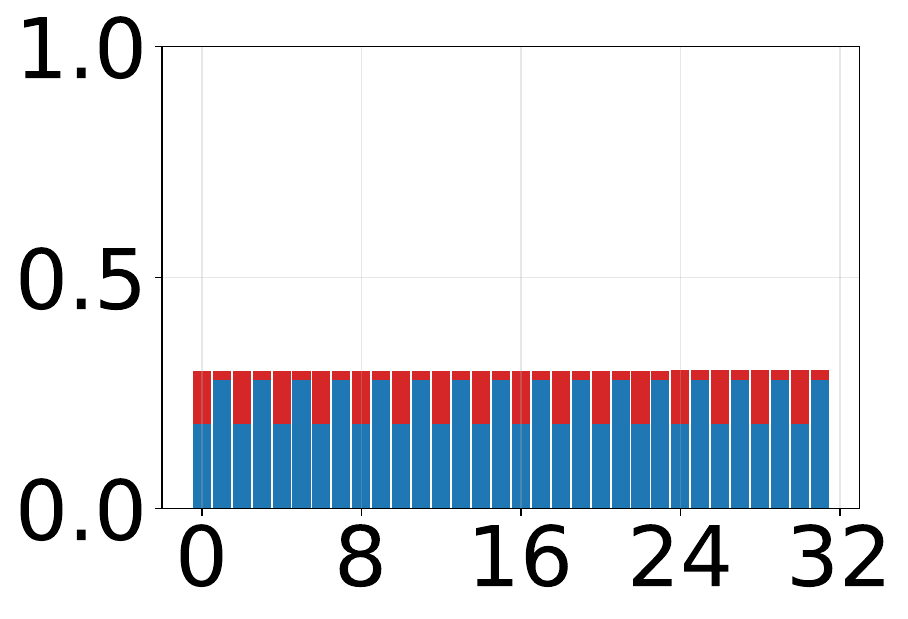}} &
    \subfloat{\includegraphics[width=0.245\columnwidth]{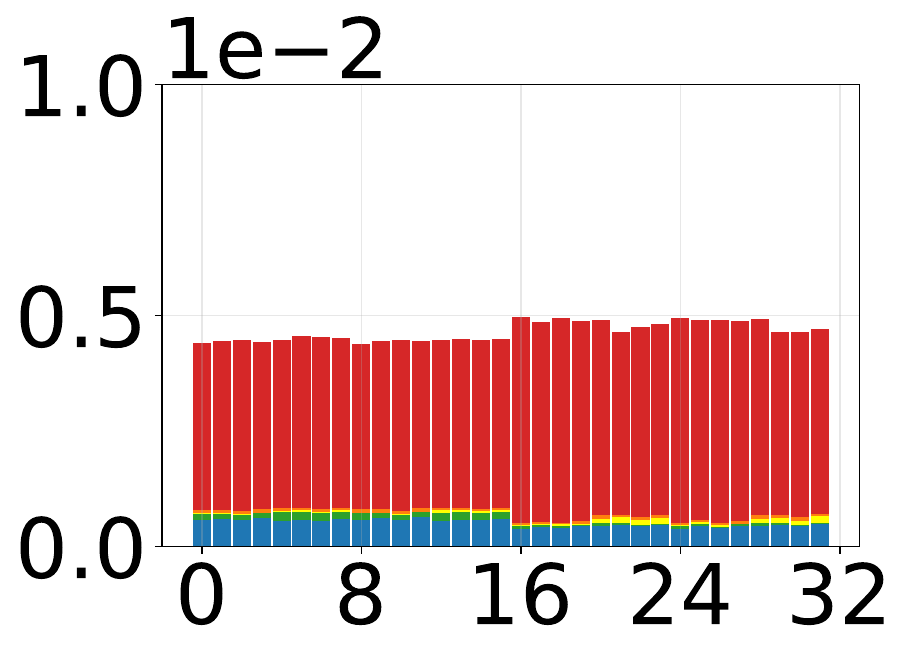}} &
    \subfloat{\includegraphics[width=0.245\columnwidth]{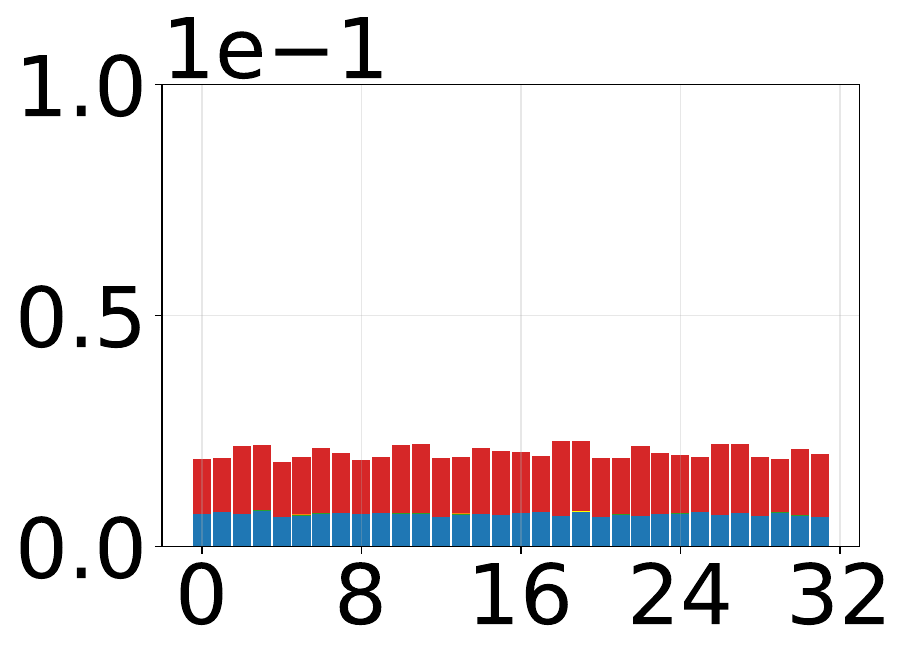}} &
    \subfloat{\includegraphics[width=0.245\columnwidth]{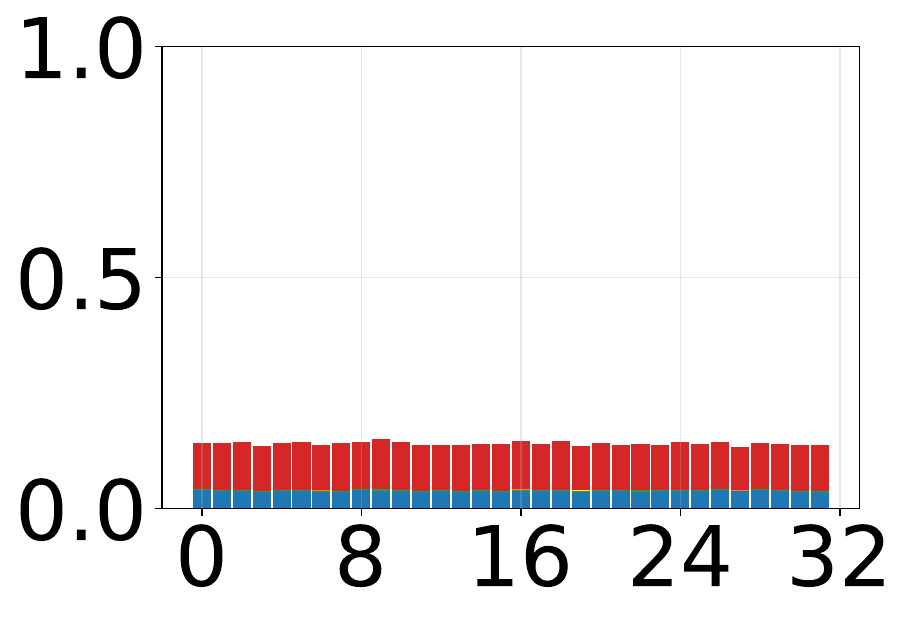}} &
    \subfloat{\includegraphics[width=0.245\columnwidth]{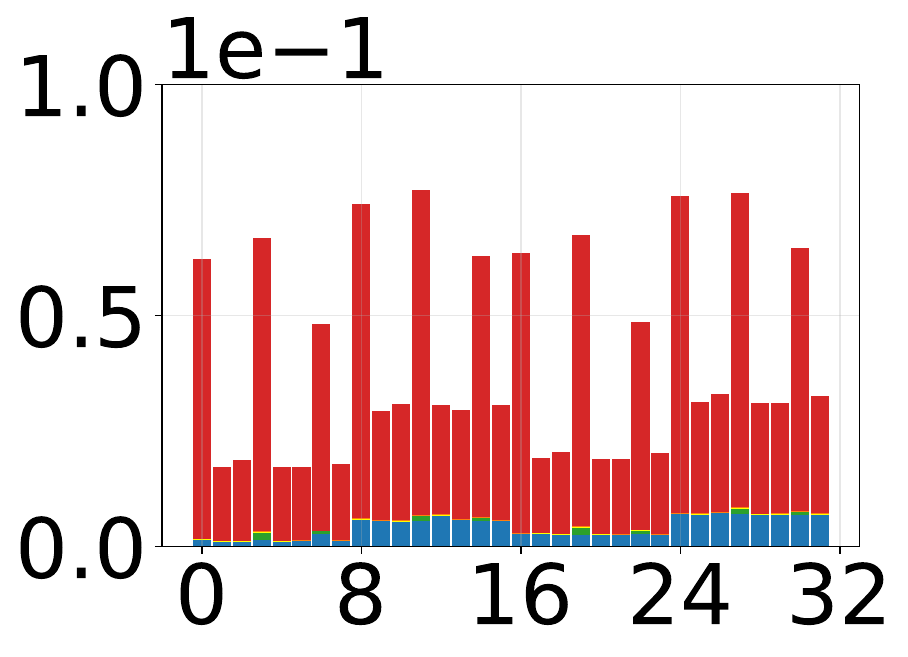}} &
    \subfloat{\includegraphics[width=0.245\columnwidth]{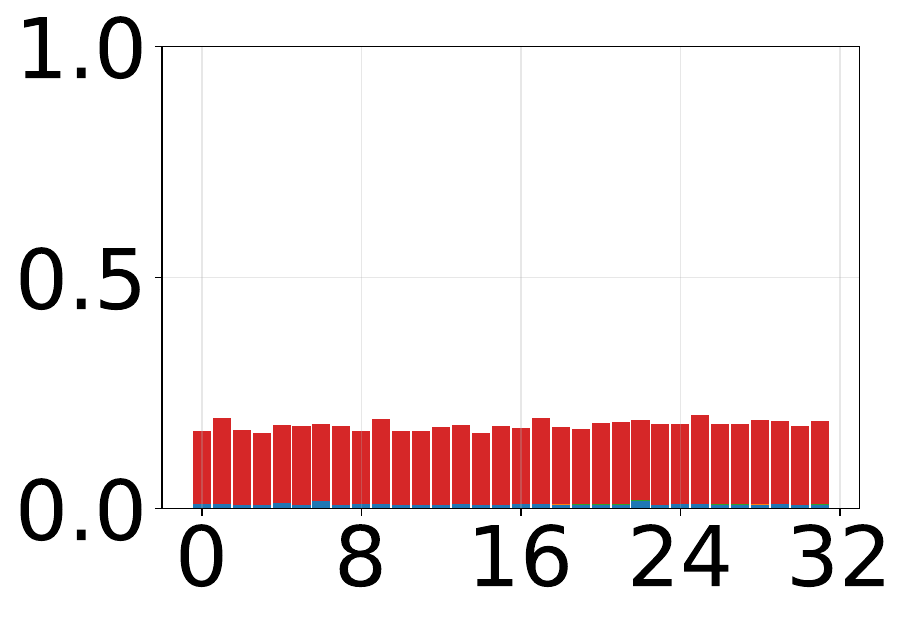}} &
    \subfloat{\includegraphics[width=0.245\columnwidth]{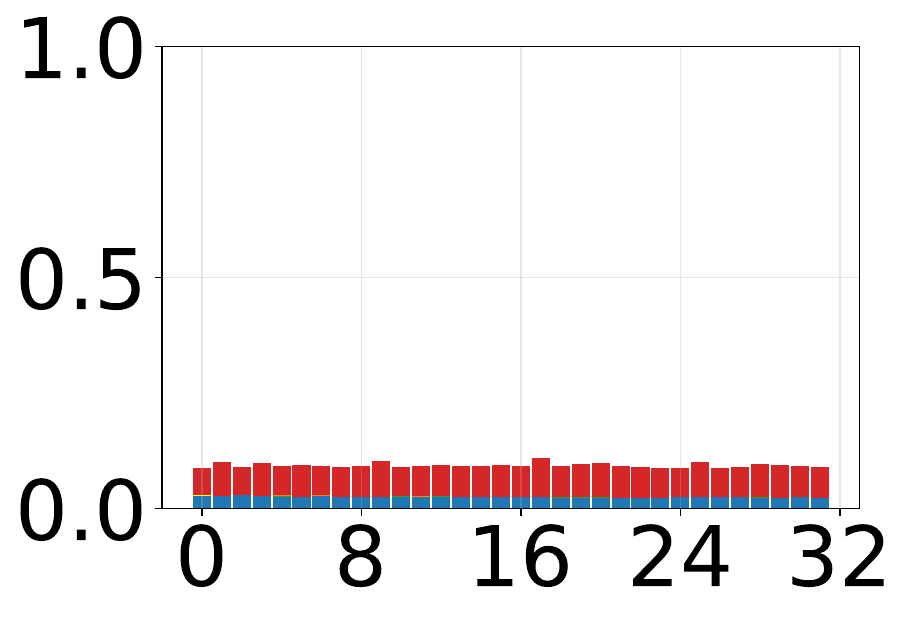}} \\
    
    \raisebox{0.65\height}{\makebox[0pt][r]{\rotatebox{90}{FP16}}} &
    \subfloat{\includegraphics[width=0.245\columnwidth]{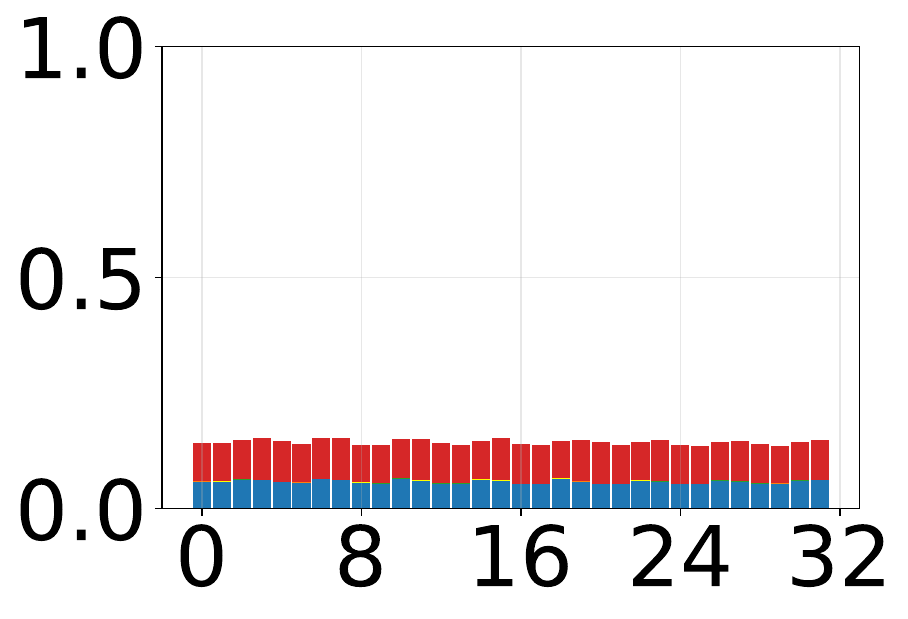}} &
    \subfloat{\includegraphics[width=0.245\columnwidth]{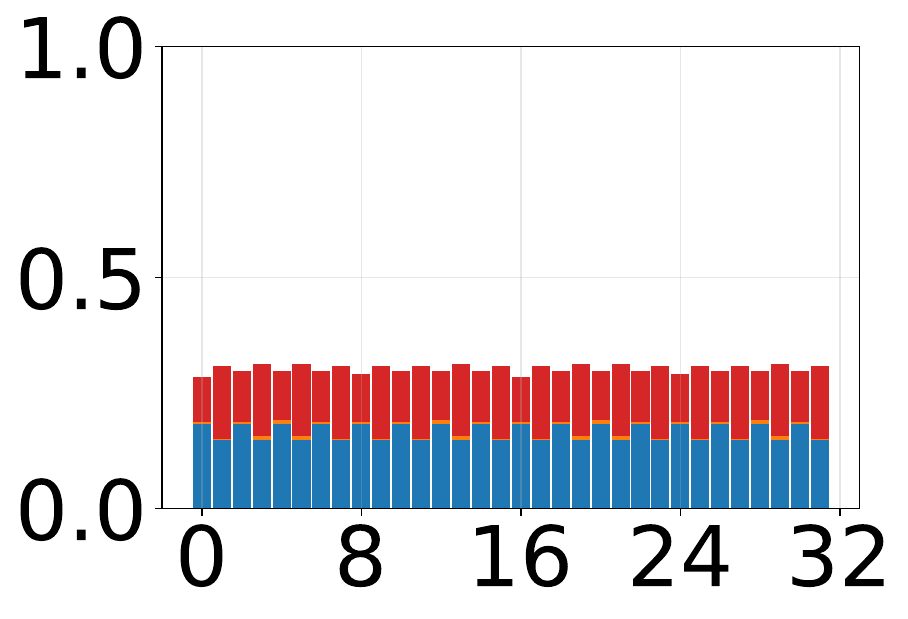}} &
    \subfloat{\includegraphics[width=0.245\columnwidth]{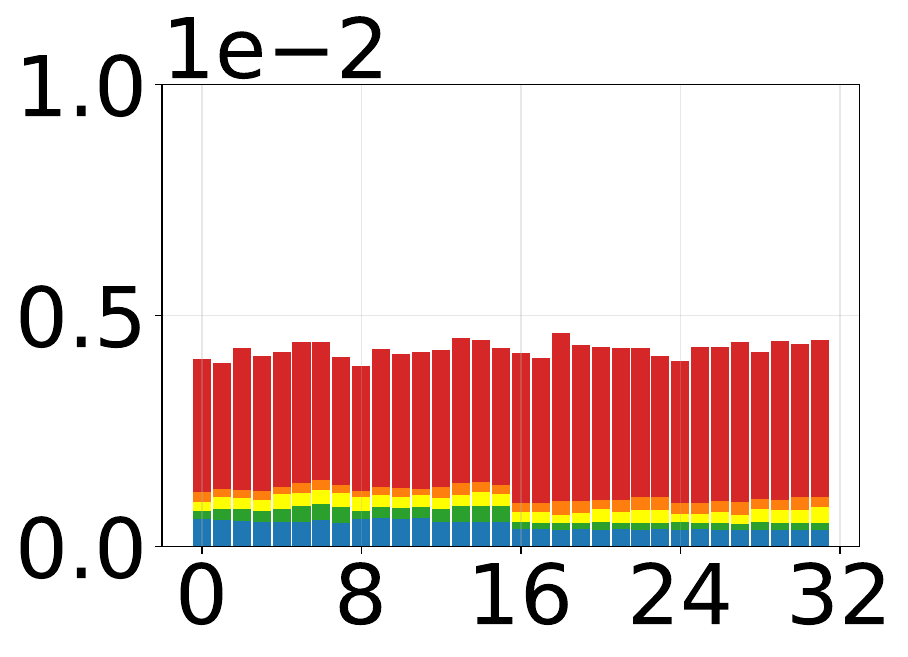}} &
    \subfloat{\includegraphics[width=0.245\columnwidth]{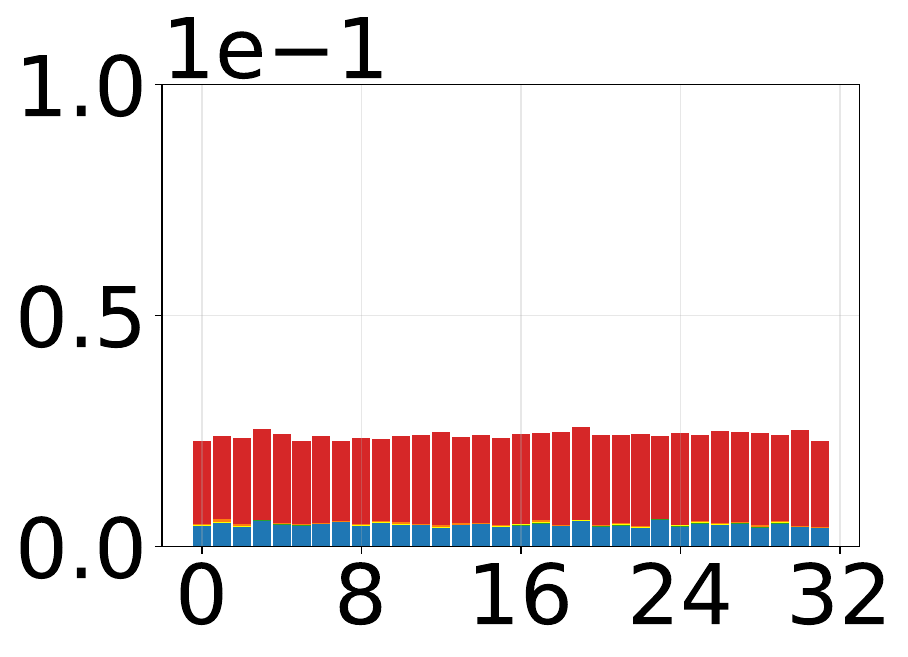}} &
    \subfloat{\includegraphics[width=0.245\columnwidth]{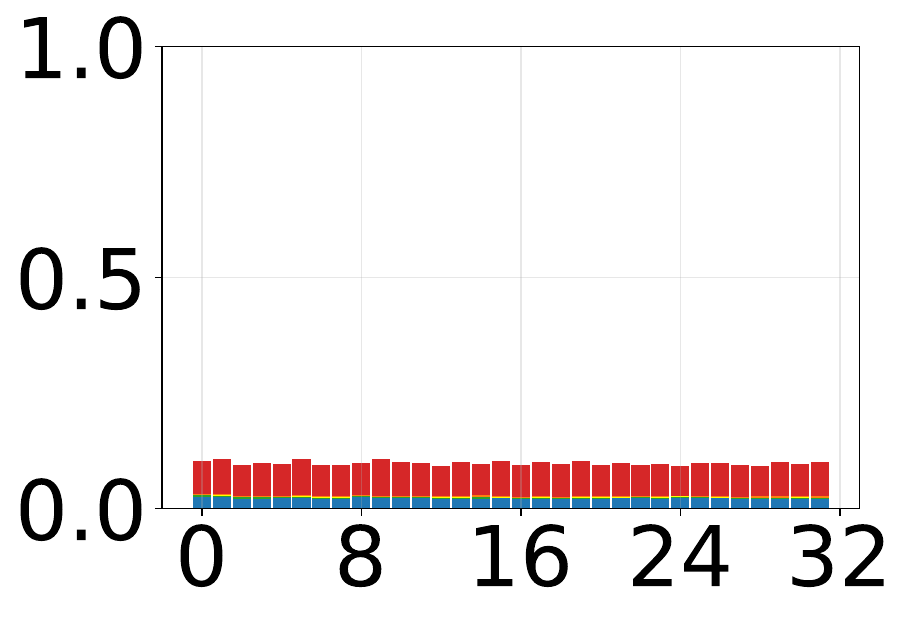}} &
    \subfloat{\includegraphics[width=0.245\columnwidth]{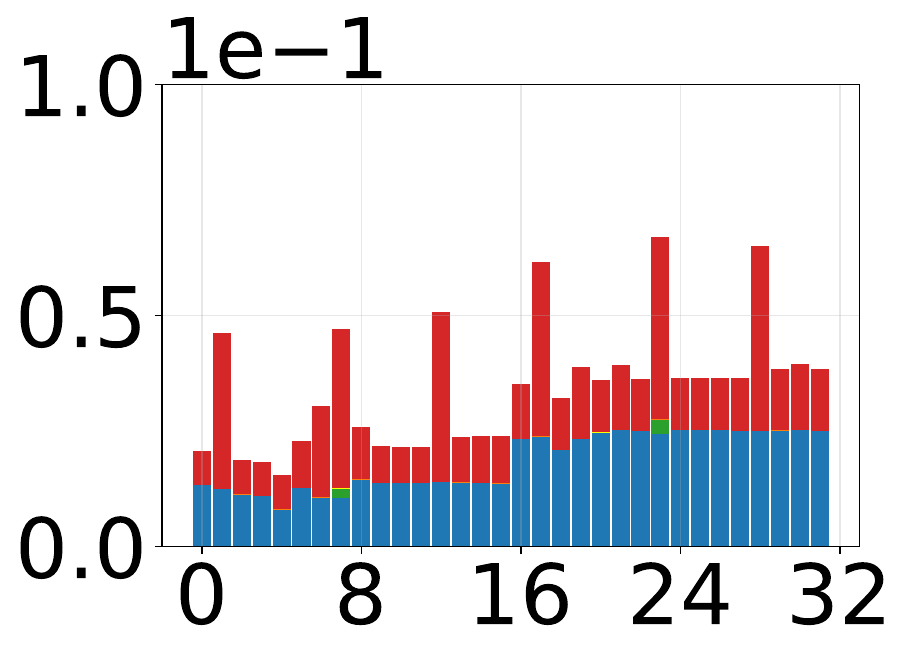}} &
    \subfloat{\includegraphics[width=0.245\columnwidth]{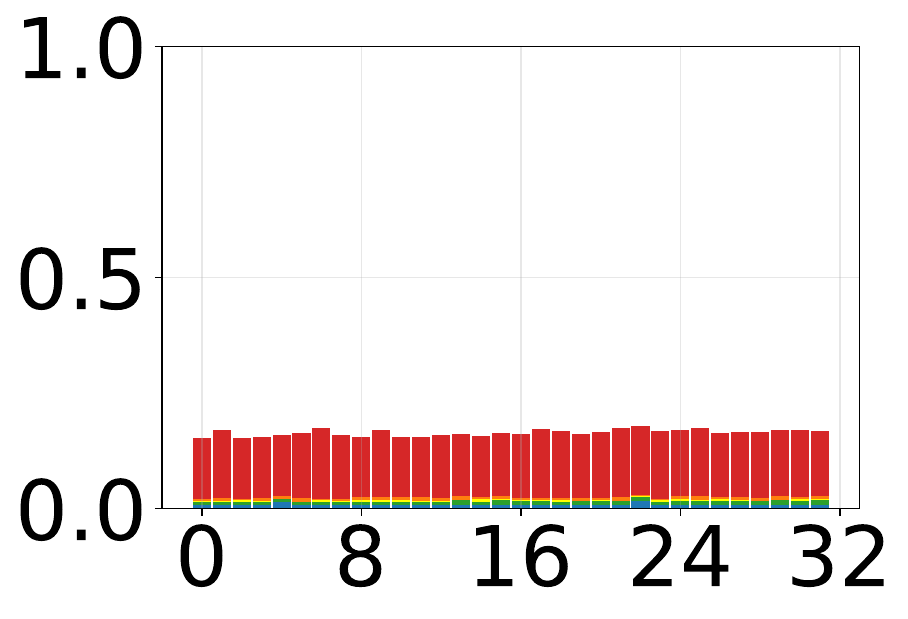}} &
    \subfloat{\includegraphics[width=0.245\columnwidth]{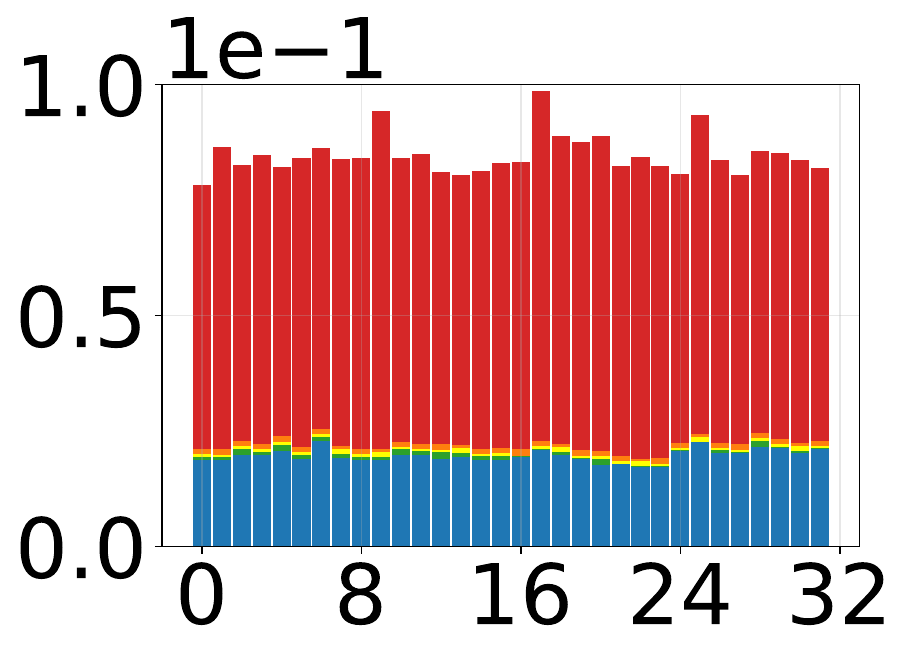}} \\
    & \footnotesize{CudaCoreControl1} & \footnotesize{CudaCoreControl2} & TensorCore & ALU
    & L1\$Data & \footnotesize{L1\$Miss Handler} & L1\$Tag & CudaCoreIO \\
  \end{tabular}
  \caption{The corruption rate (y-axis of each sub-figure) of each memory address when modulo the warp size $W$ (x-axis of each sub-figure, $W=32$). Each sub-figure represents the aggregated SDC statistics from all micro-workloads for a certain data type (listed on the left of each row) and all fault sites in a certain hardware unit (listed at the bottom of each column).}
  \label{fig: warp_stats}
  \vspace{-3mm}
\end{figure*}


As GPU execution groups threads into warps, when batched operations write to memory, a fault in one data path can manifest as multiple corruptions in strided addresses.
A common warp size is 32~\cite{gpu_programming_guide}, and our evaluated GPU model follows this design.
Accordingly, we analyze spatial correlation by grouping output addresses modulo the warp size ($W=32$).

\myobservation{SDCs exhibit spatial periodicity across hardware units, data types, and corruption categories.}

Fig.~\ref{fig: warp_stats} shows the corruption rate for each error type across all combinations of data types and hardware units (FP8 is omitted as discussed in Sec.~\ref{subsec: bitflip_stats}).
The corrupted addresses exhibit periodicity at multiple granularities, e.g.,
16 (\emph{TensorCore}: FP32/FP16; \emph{L1\$Miss}: UINT32/FP32/FP16),
8 (\emph{L1\$Miss}: UINT32/FP32/FP16; \emph{CudaCoreIO}: UINT32),
4 (\emph{ALU}: FP32), and
2 (\emph{CudaCoreControl2}: FP32/FP16).

The observed periodicity may also vary across corruption categories, in addition to fault location and workload data types; e.g., non-special corruptions repeat the pattern in a period of 8, whereas nullifications in 16 for \emph{L1\$Miss} faults.
The periodicity may arise from GPU parallelism, which groups data and threads whenever possible, leading to shared upstreams across a set of program outputs.

%% file: 6_impact.tex
\section{Implications for High-Level Modeling}

This section summarizes the key insights from our gate-level GPU FI and their implications for high-level SDC modeling.


\subsection{Revising High-Level Fault Injection Assumptions}

We distill three central insights regarding the structural characteristics of SDCs in modern GPUs in Sec.~\ref{sec: results}.


\myinsight{SDC outcomes are dominated by nullification and non-special bit-flips, with fault-location dependence.}


%
%

While prior DL accelerator studies qualitatively reported the rarity of NaN/$\pm$INF~\cite{he2023understanding}, our work on a production-class GPU quantifies their rate (1.01\%) and further reveals fault-location dependence, with control-path faults skewing toward nullification and data-buffer faults toward non-special bit-flips.
This systematic skew suggests that high-level fault injection and protection should reflect unit-specific corruption distributions rather than assuming device-wide uniform behavior.

\myinsight{Bit-flips are multi-bit and position-dependent.}


Less than 40\% of bit-flip corruptions are single-bit events, indicating that the single-bit corruption model underrepresents the observed fault behavior.
Multi-bit flip counts exhibit a broad secondary mode beyond the single-bit peak, and flip probabilities per bit position decrease from LSB to MSB, with mantissa bits more frequently affected than exponents.
These characteristics suggest that software- and micro-architecture-level FI using uniform single bit-flips~\cite{reagen2018ares, yang2024gpu, sabbagh2019evaluating, he2020fidelity} and high-level FI tools offering only single or two adjacent bit flips~\cite{tsai2021nvbitfi, chen2020tensorfi, hari2017sassifi, mahmoud2020pytorchfi} can benefit from incorporating empirically derived multi-bit and position-dependent distributions.

\myinsight{Corruptions show warp-aligned spatial periodicity.}



Corruptions exhibit structured spatial periodicity rather than random address dispersion.
High-level models assuming independent element-wise injection~\cite{yang2024gpu, he2020fidelity} may therefore misrepresent the structure of manifestations.
The periodicity could enable pattern-aware SDC predictors and fault localizers.



\subsection{A Distribution-Aware Software Injection Template}

\begin{figure}[!t]
    \includegraphics[width=0.98\columnwidth]{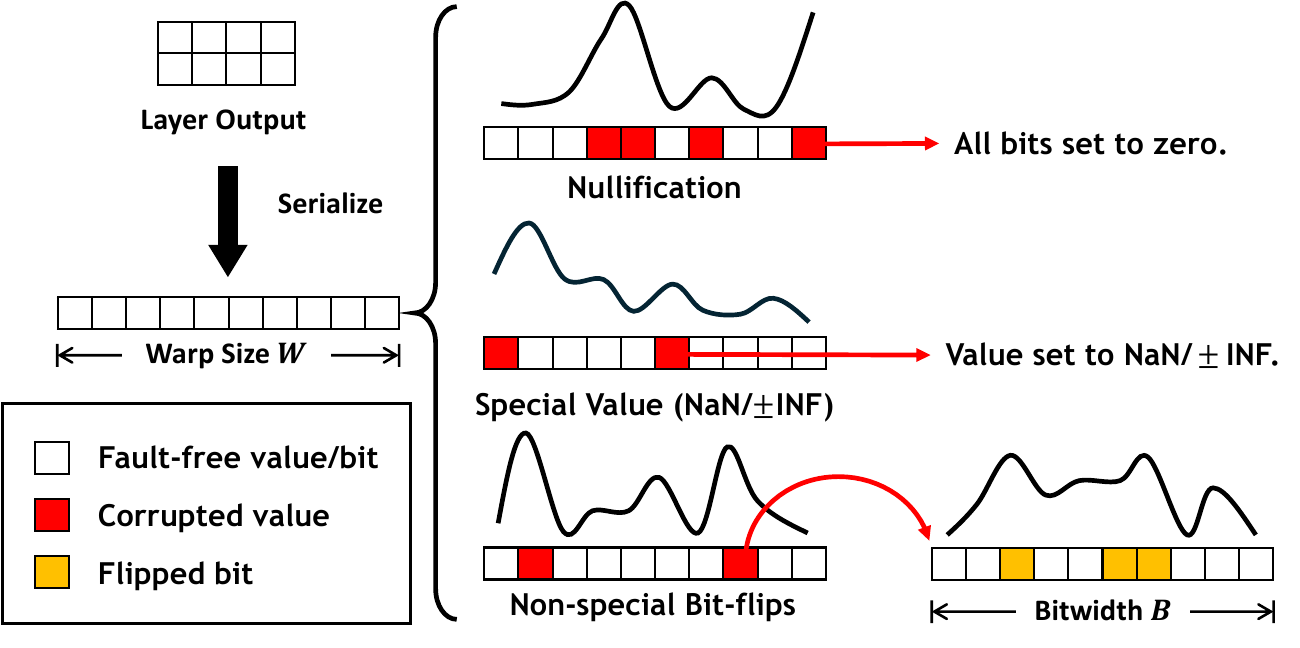}
    \caption{Software FI guided by gate-level error statistics.}
    \label{fig: sw_fi}
    \vspace{-3mm}
\end{figure}


Fig.~\ref{fig: sw_fi} translates the extracted statistics into a distribution-aware software injection template.
An output tensor is logically serialized and partitioned into address chunks aligned with the observed warp-level periodicity.
The applied error profile is conditioned on the exercised data path (kernel/data type) and the fault location (single-fault or hardware-unit level).
Within each chunk, corruption outcomes are sampled from mutually exclusive categories (nullification, NaN, $\pm$INF, or non-special bit flips), with the latter parameterized by empirically derived bit-flip distributions.
This statistical abstraction preserves the dominant outcome, bit-level, and spatial characteristics observed at the gate level while remaining computationally tractable at the software level.

%% file: 7_related_work.tex
\section{Related Work}

\myparatight{Industrial Observations of SDC.}
The presence of SDCs in large-scale data centers has been widely reported by Google~\cite{hochschild2021cores}, Meta~\cite{dixit2021silent}, and Alibaba~\cite{wang2023understanding}. 
Several major vendors, including Intel, AMD, and NVIDIA, have joined Google and Meta in the Open Compute Project (OCP) to collectively study and mitigate SDCs~\cite{ocp2024sdc}, highlighting broad industrial concern.
Recent disclosures from Amazon~\cite{ma2025understanding} and others show continued investigation of SDC effects during LLM training using faulty or aged hardware components.

\myparatight{SDC Characterization Efforts.}
Intel, Auburn University, the University of Athens~\cite{singh2023silent} identify timing marginalities as the major source of SDC in CPU.
Alibaba~\cite{wang2023understanding} collects the SDC patterns on their CPU cluster.
FIdelity~\cite{he2020fidelity} and its successor~\cite{he2023understanding} from Google propose architectural-level SDC modeling using open-source DNN accelerators (NVDLA).
Intel~\cite{bittel2024data} evaluates transient SDC propagation in ML workloads, while NVIDIA~\cite{hukerikar2024optimizing} performs large-scale FI campaigns for automotive functional safety.
%
%
Note that earlier works examined stuck-at faults propagating from flip-flops to memory~\cite{cho2013quantitative} and identified the mismatch between flip-flop-level and architectural fault injection~\cite{mirkhani2014rethinking}.
More recent analyses find that traditional FI models can miss fault scenarios leading to test escapes and silent corruptions~\cite{li2022pepr}.

\myparatight{Software-level Fault Injection and Modeling.}
SASSIFI~\cite{hari2017sassifi}, NVBitFI~\cite{tsai2021nvbitfi}, and PyTorchFI~\cite{mahmoud2020pytorchfi} frameworks enable efficient fault injection on commercial GPUs, supporting instruction- or tensor-level analysis.
These tools allow scalable evaluation of ML model resilience but rely on synthetic error models (e.g., random or single-bit flips) that lack empirical grounding.
Statistical models such as Error Model~\cite{saxena2022error} target coverage estimation rather than realistic corruption pattern prediction.

\myparatight{GPU Error Studies.}
While NCSA's Delta Project~\cite{cui2025characterizing} characterizes DUEs and their impact in GPU clusters, studies on GPU SDC remain scarce.
Prior GPU FI works typically target control or scheduler logic via instruction-level injectors~\cite{wei2023approxdup, guerrero2023understanding}, limited to specific kernel types or CNN workloads.
As a result, comprehensive gate-level SDC characterization on modern data center GPUs remains underexplored.

%% file: 8_conclusion.tex
\section{Conclusion}

To characterize silent data corruption (SDC) on a production-class GPU, we conducted a large-scale gate-level fault injection campaign, consuming over three million simulator hours across a comprehensive CUDA micro-benchmark suite.
Across hardware units, fault sites, and data types, we quantify corruption-type ratios (NaN, $\pm$INF, nullification, and bit-flips) and characterize bit-flip distributions in which single-bit flips constitute a minority.
We further identify warp-aligned spatial periodicity at multiple scales.
These findings indicate that GPU SDC behavior is structured rather than independent, motivating higher-level models that capture bit-level and spatial dependencies.
This work establishes a quantitative basis for future GPU reliability analysis and SDC mitigation.